# Exploiting Challenges of Sub-20 nm CMOS for Affordable Technology Scaling

Submitted in partial fulfillment of the requirements for

the degree of

Doctor of Philosophy

in

Electrical & Computer Engineering

Kaushik Vaidyanathan

B.E., Electronics & Communication Engineering, Anna University
M.S., Electrical & Computer Engineering, Carnegie Mellon University

Carnegie Mellon University
Pittsburgh, PA

March, 2014

*To my family, teachers and friends*

# Abstract

For the past four decades, cost and features have driven CMOS scaling. Severe lithography and material limitations seen below the 20 nm node, however, are challenging the fundamental premise of affordable CMOS scaling. Just continuing to co-optimize leaf cell circuit and layout designs with process technology does not enable us to exploit the challenges of a sub-20 nm CMOS. For affordable scaling it is imperative to work past sub-20 nm technology impediments while exploiting its features. To this end, we propose to broaden the scope of design technology co-optimization (DTCO) to be more holistic by including micro-architecture design and CAD, along with circuits, layout and process technology. Applying such holistic DTCO to the most significant block in a system-on-chip (SoC), embedded memory, we can synthesize smarter and efficient embedded memory blocks that are customized to application needs.

To evaluate the efficacy of the proposed holistic DTCO process, we designed, fabricated and tested several design experiments in a state-of-the-art IBM 14SOI process. DTCO'ed leaf cells, standard cells and SRAM bitcells were robust during testing, but failed to meet node to node area scaling requirements. Holistic DTCO, when applied to a widely used parallel access SRAM sub-block, consumed 25% less area with a 50% better performance per watt compared to a traditional implementation using compiled SRAM blocks and standard cells. To extend the benefits of holistic DTCO to other embedded memory intensive sub-blocks in SoCs, we developed a readily customizable smart memory synthesis framework (SMSF). We believe that such an approach is important to establish an affordable path for sub-20 nm scaling.

# Acknowledgements

First, I would like to thank my advisor Professor Larry Pileggi who through his continued guidance and support has helped me persevere through graduate school. Larry provided me with many opportunities that have helped me evolve as a researcher. Without his vision and technical contributions this interdisciplinary and collaborative research would not have been possible. To me, he is someone I always look up to.

Next, I would like to thank Dr. Lars Liebmann at IBM for mentoring me over the past several years. Several key ideas in this work were born and took shape during those weekly interactions with Lars. This work has been largely possible owing to his foresight and managerial acumen. I am grateful to Professor Andrzej Strojwas for his constant encouragement and insightful discussions. I would like to acknowledge Professor Shawn Blanton for taking time out to be on my thesis committee and providing me with valuable feedback.

My first few years in graduate school would have been difficult if not for Daniel Morris. Starting from those tapeouts in my very first month in graduate school, Dan has always been around to teach, guide and help. Several pieces of my early research, such as construct-based design, were done in collaboration with him. Qiuling (Jolin) Zhu has always impressed me with her work ethic. Collaborating with her on the smart memory synthesis framework was a rewarding experience. I would also like to thank Prof. Franz Franchetti for his help with identifying applications for smart memories, and Prof. Mark Horowitz at Stanford for lending the Genesis framework. I would like to thank Ekin Sumbul, Mitchell Bender, Renzhi Liu, Siew Hoon NG, Wenbin Huang, Bishnu P Das and Prashant Kashinkunti for contributing to various aspects of this work. Special thanks to David Bromberg for constantly inspiring me with his terrific work ethic. I am also grateful to Adam Palko, Samantha Goldstein, Charlotte Ambrass, Judy Bandola and Elaine Lawrence for helping me navigate through administrative aspects in the department.

A significant portion of this work was done in collaboration with IBM. I would like to acknowledge Kafai Lai, Michael Guillorn, Neal Lafferty, Jassem Abdallah and Matthew Colburn for help with lithography aspects of this work; Steve Wu and Mark Liu for wafer testing; Leon Sigal and Greg Northrop for sharing their considerable design and technology expertise.

This work was sponsored by the DARPA GRATE (Gratings of Regular Arrays and Trim Exposures) program under Air Force Research Laboratory (AFRL) contract FA8650-10-C-7038. I would like to acknowledge DARPA for funding the GRATE program as it helped create and nourish the collaborative ecosystem required to undertake such interdisciplinary research.

I am grateful to Circuit Research Labs in Intel Corporation for offering me an opportunity to work with Ram Krishnamurthy, Amit Agarwal and Steve Hsu as a research intern. Working at Intel Labs was a great learning experience providing me with a much-needed orientation to research in an industrial setting.

I would fail in my duty if I do not thank Professor Ken Mai, Professor Rob Rutenbar and Professor Wojciech Maly for introducing me to the domains of integrated circuits, CAD and manufacturing, respectively. I would like to acknowledge my professors in undergrad and teachers in school for instilling in me a curiosity for electronics and science.

Graduate school would have not been enjoyable without the companionship of my friends Arun, Bhadri, Anusha, Ashwati, Abhay, Kaustubh, Varoon, Sudharsan, Mukund, Vivek, Kedar, Anirban, Abhinav, and many others.

Last, but not the least, I would like to thank my parents, grandmother, sister, brother, uncle, aunt and all my relatives for helping me grow as an individual.

Finally, graduate school would have been impossible without my best friend and wife Lavanya. I will forever cherish graduate school for the memories it has gifted us.

Table of Contents

# List of Figures

# List of Tables

# 1 Introduction

The well-known Moore's law [1] and Dennard's scaling theory [2] laid the foundation for cost-effective and efficient integrated circuit miniaturization. This unprecedented capability to miniaturize, thereby decreasing cost per function while still improving system efficiency indeed has distinguished electronics, and specifically CMOS, in comparison to other implementation substrates. Continuing on the CMOS technology scaling roadmap is not just essential for advancing science and technology, but also for enabling the world's socio-economic growth.

For the past decade the semiconductor industry has been firefighting several issues, mainly pertaining to lithography and material limitations, that have challenged affordable CMOS technology scaling. While there are several proposed alternatives to CMOS, none have proved to be technologically and economically viable to supplant CMOS as yet. As CMOS technology scaling continues, the 14 nm node has been seen as an inflection point for two reasons. First, the industry is adopting FinFETs and multiple patterning for critical mask levels; this move is dubbed as the most significant move the industry has made since the adoption of CMOS [3]. The second reason is that escalating complexity and cost is threatening affordable technology scaling, especially for medium and low volume semiconductor manufacturers [4]. These challenges require us to explore design and CAD techniques to drive cost-effective scaling into the future.

Standard cells and embedded memory bitcells form the most fundamental building blocks of a digital SoC. Iteratively co-optimizing the two leaf cell layouts with process technology, while enabling us to tradeoff manufacturing complexity with design efficiency, proves to be inadequate to meet node to node area scaling requirements. This raises serious concerns as fundamental leaf cells, especially embedded memory (SRAM) bitcells, have historically driven SoC scaling. To work past modern CMOS technology's impediments while exploiting its features, we propose to broaden the scope of conventional design technology co-optimization to include micro-architecture design, circuits and CAD, along with layout and process technology.

Instead of conventionally compiling SRAM blocks from inefficient hard IP leaf cells, we synthesize smart SRAMs tailored to application needs from a library of augmented bitcell arrays and standard cells. Application-specific customization guarantees efficiency while synthesis from robust components ensures productivity and manufacturability of the smart SRAM blocks. Test chips designed, fabricated and tested in a state-of-the-art IBM 14SOI process demonstrate the efficacy of the proposed design and CAD methods to drive affordable scaling.

This document is arranged as follows:

- Chapter 2 discusses different challenges impeding cost-effective CMOS scaling below the 20 nm node. It also surveys current approaches to scaling while describing how they fail to enable affordable scaling in the future.
- Chapter 3 describes our design technology co-optimization effort for two fundamental leaf cells, namely, SRAM bitcells and standard cells. This exercise suggests that leaf cell DTCO is insufficient to guarantee manufacturability and design efficiency below the 20 nm node.
- Chapter 4 details our approach to broaden DTCO to include micro-architecture, circuit design and CAD to build efficient SoCs. Specifically, we outline our approach to synthesize application-specific embedded memory blocks that showcase the benefits of a holistic DTCO process.

- <u>Chapter 5</u> presents results from our IBM 14SOI test chips containing several experiments to evaluate the efficacy of the proposed design and CAD approaches.

- <u>Chapter 6</u> concludes with a discussion of key results and proposals for future work.

# 2 Background and Motivation

In this chapter we will present the challenge of cost-effective system-on-chip (SoC) scaling below 20 nm CMOS. We will describe some prominent design approaches and show that they are inadequate to enable affordable scaling.

## 2.1 SoC Scaling Challenge

Cost and features have historically driven scaling, a result of the prophetic observation made by Gordon Moore [1]. Miniaturization or scaling has been the backbone for the electronics revolution that has greatly enhanced the quality of our lives. To put the impact of scaling in perspective, if the automotive industry had achieved similar improvements as the electronics industry, a Rolls-Royce would cost only $40, could circle the globe eight times on a gallon of gas with a speed of 2.4 million mph [5]. This promise of scaling has attracted heavy investment of resources, both financial and human, for over four decades.

By the end of the twentieth century, after serving the industry well over two decades, Dennard's classical constant field scaling ended [2]. The golden era of shrinking all dimensions of the transistor to decrease latency, power and area was over. While many anticipated this to be the end of the electronics revolution, researchers adopted new materials and process integration schemes to keep scaling alive [6]. A cursory look at specifications of a few recent microprocessors from Intel and SoCs from Apple are sufficient to convince us that consumer needs and market compulsions continue to drive scaling [7] (Figure 2.1). However, the path forward is fraught, with tough technology, design and CAD challenges.

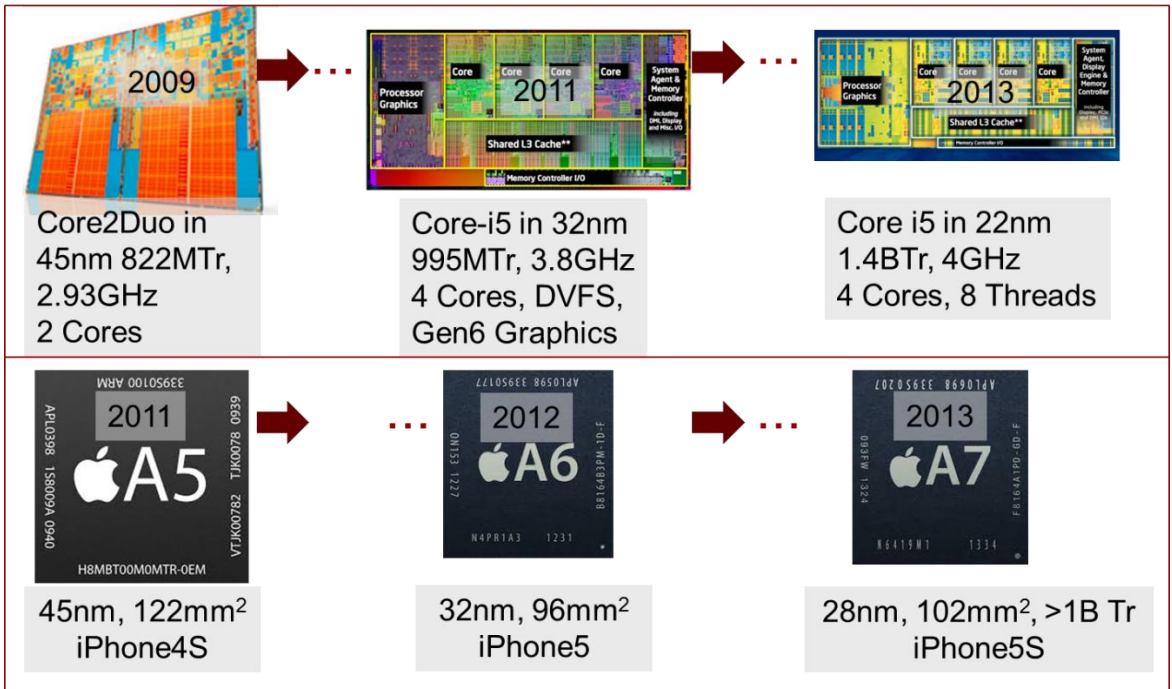

**Figure 2.1 Post-Dennard microprocessor and SoC scaling.**

## 2.1.1 Technology Challenges

As we have continued to scale in the previous decade, wafer prices have been gradually rising owing to the growing complexity of nanoscale CMOS processes. Following conventional practices, however, a recent chart from Nvidia shows how wafer price is expected to dramatically increase below the 20 nm node, thereby challenging the very premise of Moore's Law scaling (Figure 2.2) [8]. Furthermore, the huge non-recurring expense (NRE) to invest in 22 nm fabrication equipment and process development is expected to exceed 10 billion dollars, which is well beyond what many semiconductor industries can invest (Figure 2.3) [5].

Lithographic pitch scaling has driven CMOS scaling. As the industry, in the absence of the much anticipated extreme ultra-violet lithography (EUV-L), has been stuck with 193 nm immersion (193i) lithography for over a decade, lithographers have had to use several resolution enhancement techniques (RET) to achieve full pitch scaling (scaling pitch node to node by a factor of 0.7). RETs, by design, increase complexity and cost to achieve better resolution. As we scale past the 20 nm node, however, full pitch scaling is becoming unaffordable, threatening the 50% node to node area scaling (Figure 2.4).

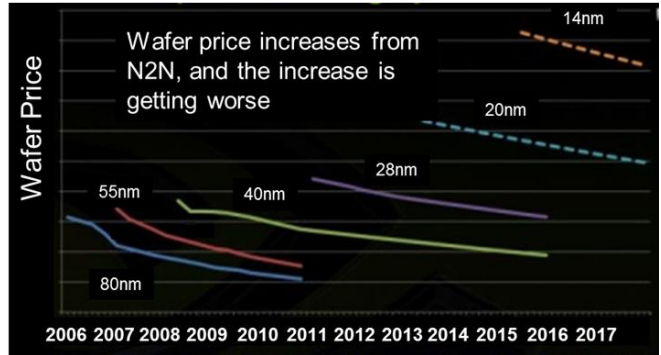

**Figure 2.2 Wafer price projected to increase with scaling [8].**

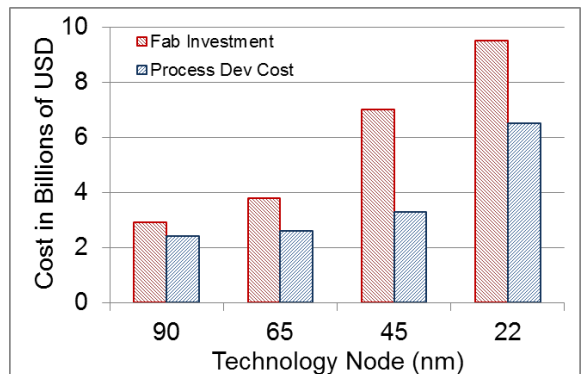

**Figure 2.3 Manufacturing cost increasing dramatically below 45 nm [5].**

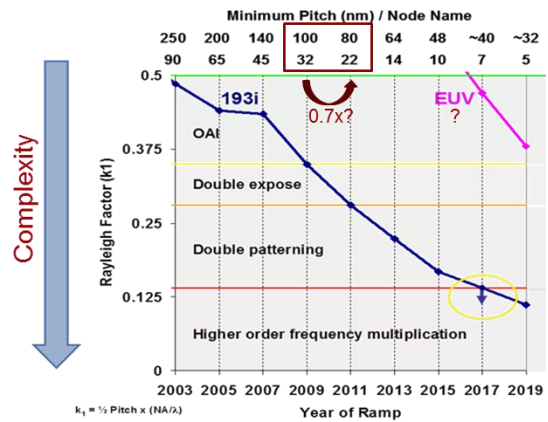

**Figure 2.4 Lithographic pitch scaling drives technology scaling. Courtesy - Dr. Lars Liebmann, IBM.**

Apart from CMOS process complexity and cost, the planar device has been a source of serious concern owing to high variability and leakage [9]. The poor gate control of the channel is being remedied temporarily using incremental improvements in materials and structures. However, to have a viable device that can scale below the 20 nm node, the industry is moving from planar CMOS to FinFETs. FinFET technology has the gate surround the channel on two or more sides, thereby providing better gate control over the channel [10]. Translating these FinFET benefits from simulation to high-volume silicon remains a challenge.

Interconnect scaling, some researchers claim, presents a more severe concern than device scaling [11]. Interconnect pitch scaling below the 20 nm node is expected to see a dramatic increase in wire resistance, thereby resulting in an increase in interconnect delay. The decreasing widths also increase the likelihood of electromigration, a major reliability challenge. Warnock et al. provide a more detailed description of interconnect scaling challenges anticipated at the 14 nm node [11].

### 2.1.2 Design Challenges

The technology challenges described in the previous section are permeating up to significantly impact SoC design. Owing to technology restrictions, specifically patterning, the two fundamental building blocks of an SoC are not scaling in a cost effective manner. Recent data presented at ISSCC trends 2013 shows sub-optimal bitcell area scaling [12]. Figure 2.5 shows that at the 20 nm technology node, the ideal 50% node to node area scaling is not seen for heavily design-technology co-optimized bitcells. Such a trend raises significant concerns as bitcells form the basic building block of SRAMs that occupy more than 50%-80% of a modern SoC area. An equally alarming trend can be observed for standard cell logic as shown in Figure 2.6, with an increase in cost per gate below the 20 nm node [13]. Both standard cell and bitcell trends clearly illustrate that the 14 nm node could be the inflection point where the industry would be forced to make several changes in design and manufacturing to keep scaling alive and profitable.

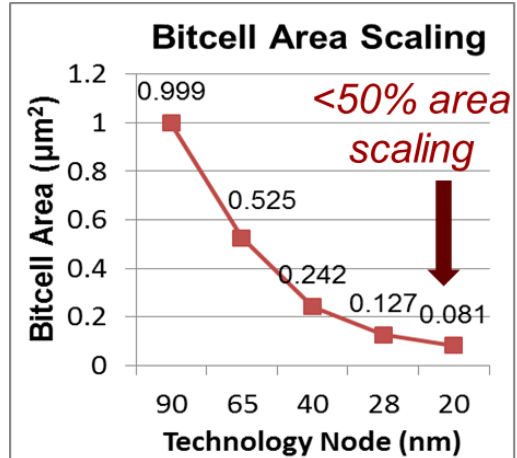

Figure 2.5 Sub-optimal bitcell area scaling [12].

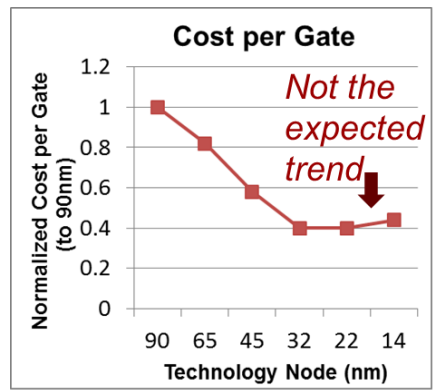

Figure 2.6 Cost-per-gate increasing as we scale below 22 nm. Courtesy - Dr. Subbu Iyer, IBM.

Unforgiving to these challenges is the ever increasing consumer demand for more feature rich and high performing electronic gadgets. This is requiring IC design companies to pack more and more functions on to an SoC, dramatically increasing design complexity. For instance, Apple's A7 SoC driving the iPhone5S has more than1 billion transistors [7]. Projections from TSMC (Taiwan Semiconductor Manufacturing Company) indicate that SoC design at the 16 nm node would be at least three times more complex than a design at the 28 nm node, with significant time spent on hard IP (intellectual property) qualification, physical design and system verification, as compared to actual architecture design [14] (Figure 2.7).

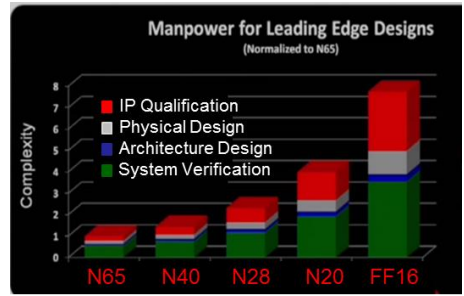

**Figure 2.7 Dramatic increase in design complexity (turnaround time) below the 28 nm node [14].**

### 2.1.3 CAD Challenges

CAD tools and methodologies have enabled us to effectively and robustly design SoCs with more than a billion transistors in less than year. However, Semico Research projects a dramatic increase in the complexity of CAD tools and design flows as we scale to the 28 nm node and beyond, indicated by the 102% increase in CAD tool cost [15]. Specifically, owing to increased process complexity, physical verification methodologies are becoming more involved. Auto-routers, used for automatically connecting different gates in the SoC, are becoming more resource and run time intensive [16], indicated by the increasing number of lines in a router technology file (Figure 2.8). Furthermore, with the use of double patterning in the 14 nm node, physical design CAD tools and flows are expected to become more complex and design-time intensive, posing new challenges for CAD tool developers and designers.

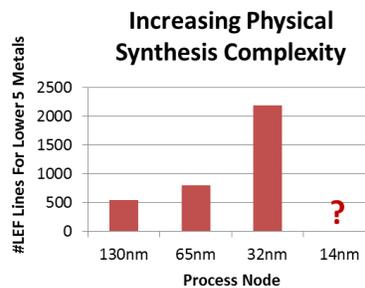

**Figure 2.8 Increasing auto-router technology file complexity with technology scaling.**

Zvi Or-Bach also summarizes the challenges of affordable scaling eloquently in [17]. At this juncture it is crucial to revisit if conventional design and CAD techniques can cope up with the mounting challenges in technology, design and CAD.

## 2.2  Scaling at Sub-20 nm CMOS

Planar CMOS has driven the semiconductor industry for more than three decades. Owing to excessive leakage and variation, the industry is moving from planar CMOS to FinFET below the 20 nm node, dubbed as the most significant move by the industry in many decades [3]. Figure 2.9 (a) shows a 3-fin FinFET. FinFET has the gate covering the channel (fins) on more than one side, providing better gate control over the channel. As it might be apparent to a designer, FinFETs only allow discrete channel widths as compared to continuous channel widths provided by planar CMOS. FinFETs are expected to scale to at least the 7 nm node [18].

The process stack of the FinFET device is shown in Figure 2.10 (a) and compared with a planar CMOS process stack (Figure 2.10 (b)). FinFET devices use a long rectangular contact, also called a local interconnect, to strap multiple fins together to form a multi-fin device. Local interconnects provide connectivity to the extremely restricted fin and poly layer, and is another new technology element in a sub-20 nm CMOS process [37].

With FinFET devices, the next major technology element that the industry would be forced to adopt is multiple patterning. As pitches of critical layers, namely device and 1X metal layers, scale below the maximum resolution achievable with single patterning using 193i lithography, the industry has resorted to using multiple mask layers to pattern one design layer. While multiple patterning significantly increases manufacturing cost, it also poses several challenges for designers and CAD developers. A scanning electron microscope (SEM) image of Metal1 (M1) layer that is manufactured using three mask levels is shown in Figure 2.9 (b).

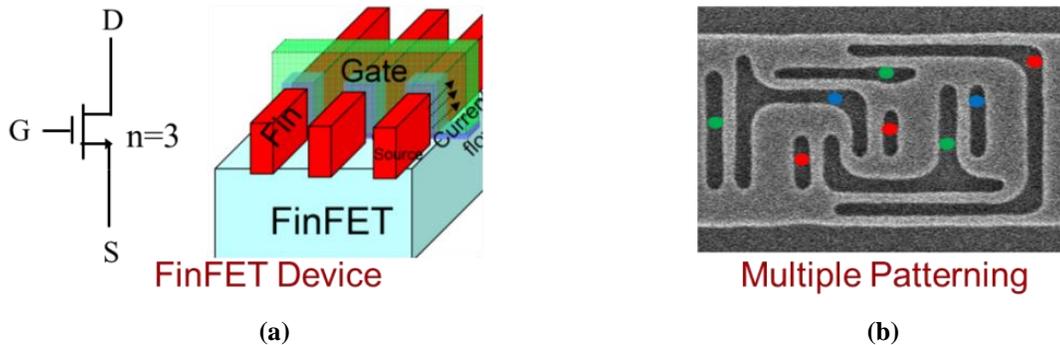

Figure 2.9 Differentiating elements in a 14 nm process (a) FinFET device [19], (b) Multiple patterning.

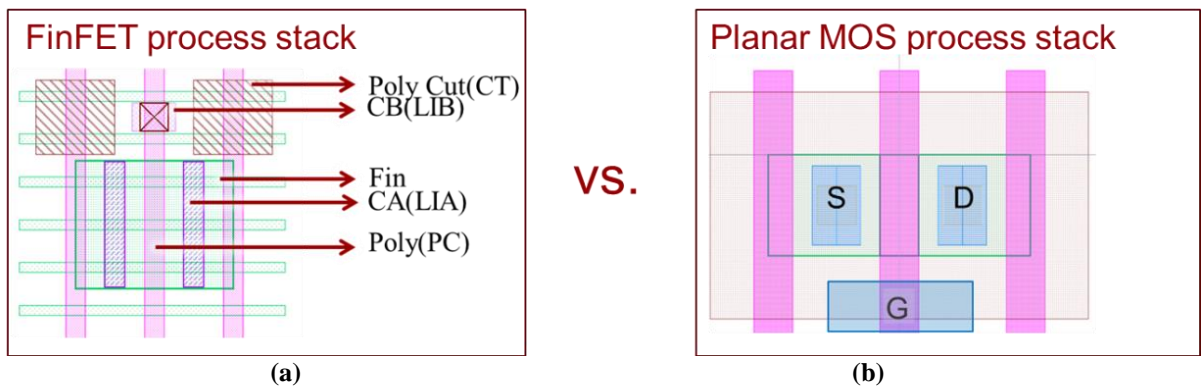

Figure 2.10 Process Stack: (a) FinFET, (b) Planar CMOS.

## 2.2.1 Sub-20 nm Technology Offering

In this section we provide a detailed description of a 14 nm FinFET device using an example NAND2_X1 gate (Figure 2.11 (a)). FinFET devices use a grating based fin layer as compared to the more complex active region shapes seen at previous nodes. This process uses unidirectional and gridded local interconnect layers to contact gate (CB) and active area (CA). The contact layer (V0) connects the local interconnect (CA and CB) to Metal 1 (M1). The FEOL-layer and BEOL-layer connectivity is illustrated using a NAND2_X1 layout in Figure 2.11 (b) and Figure 2.11 (c) respectively. One important observation is that the FEOL layers are unidirectional and gridded; i.e., one-dimensional (1D) gratings. While FEOL layers are unquestionably grating-based, the design-manufacturability tradeoffs for the BEOL layers are less agreed upon. There is a strong incentive to make M1 and other 1X BEOL layers grating-like as it lessens hot-spot risk by limiting the number of unique layout neighborhoods and reduces the number of exposures, simplifying layout decomposition. However, a more restricted BEOL layer curbs design freedom, potentially resulting in an inefficient design. Therefore, agreeing on a good technology definition requires consideration of design and technology needs.

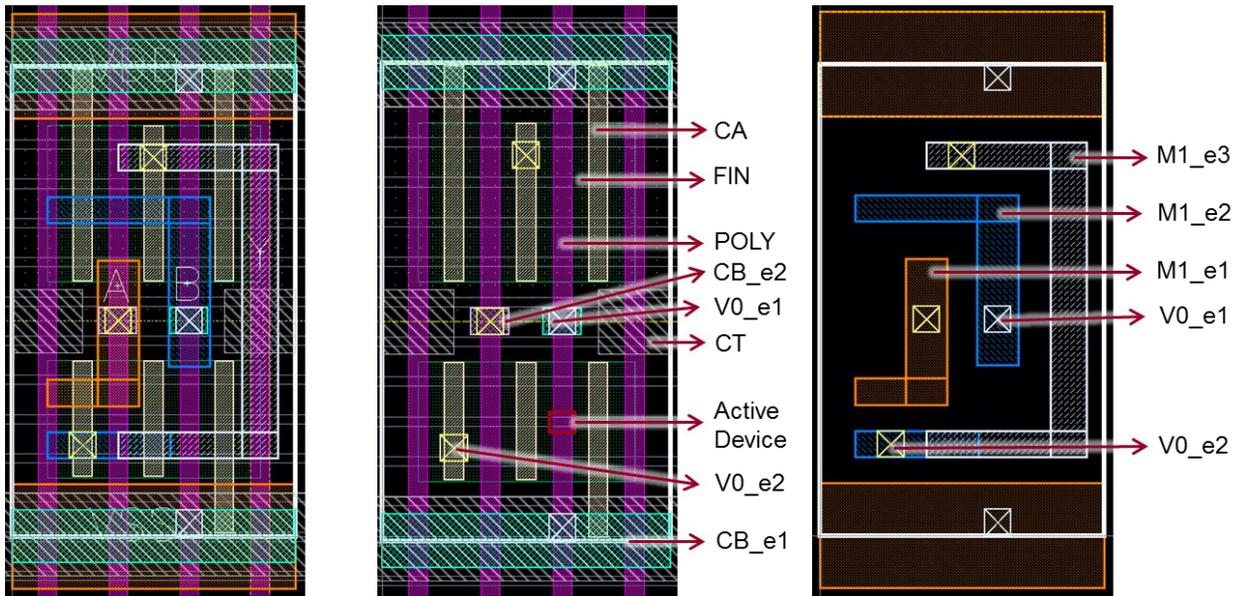

**Figure 2.11 (a) 14nm NAND2_X1 layout; (b) FEOL + Local Interconnect + V0 of NAND2_X1; (c) BEOL + V0 of NAND2_X1.**

### 2.2.2 Conventional Design Scaling Approaches

In this section we assess the effectiveness of a few prominent design scaling approaches to tackle sub-20 nm SoC design and manufacturing challenges.

#### 2.2.2.1 *Geometric Shrink*

One classic technique to scale leaf cells, especially standard cells, is to keep the same layout topology as in the previous technology nodes, while meeting design rules in the current node. As this technique has been very productive in the previous nodes, we attempted a geometric shrink of a 32 nm NAND gate to meet 14 nm design rules. As shown in Figure 2.12, the resulting NAND gate is area inefficient, has non-connectable inputs pins, requiring at least triple patterned M1 layer and has several minimum M1 area landing pads. Clearly, geometric shrink is not a technique to scale cost-effectively.

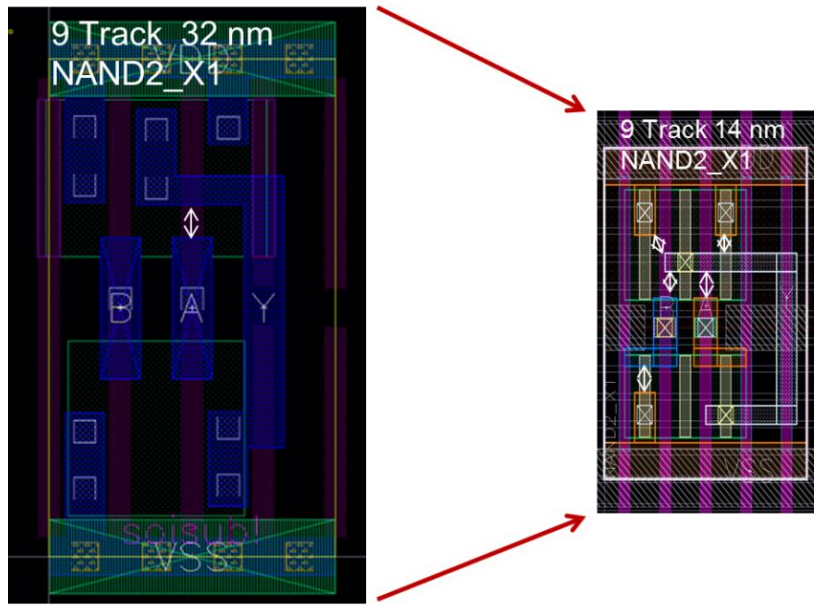

**Figure 2.12 Geometric Shrink from 32nm layout to 14nm node.**

### 2.2.2.2 *Extremely Restricted Design Rules*

As the number of design rule escapes dramatically increased causing yields to plummet, the industry adopted restricted design rules (RDR) [20][21]. RDRs tradeoff design freedom for robust manufacturability and have met with reasonable success in 45 nm and 32 nm nodes. Few researchers have proposed extreme RDRs for heavily patterning restricted nodes below 20 nm [32][35]. Specifically, they propose to adopt layouts from previous technology nodes and then replace previously used bidirectional layout shapes with two unidirectional layout shapes and a via. To test the effectiveness of this approach, we build a NAND gate by keeping the same layout topology as the previous technology nodes, but replacing bidirectional M1 with a vertical

M1, horizontal Metal2 (M2) and via1 (V1). The resulting gate is area inefficient, increasing the number of vias and the number of process sensitive minimum M1 area landing pads. This exercise illustrates that such a localized approach does not solve, but will only move the problem from patterning to process integration.

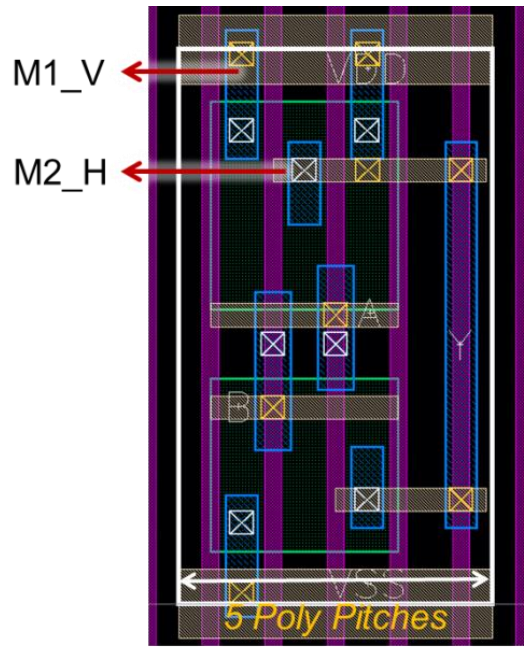

Figure 2.13 Creating 1D-M1-M2 layout from 2D-M1 layout.

### 2.2.3 Design Technology Co-Optimization Enables Affordable Scaling

Conventional design scaling techniques are not adequate to address the complex challenges posed by nanoscale CMOS and drive affordable scaling [22]. Well aware of their limitations, the industry is moving away from forcing designers to build components from a predefined process. On the contrary, a significant thrust has been toward providing a design-aware technology definition. Design technology co-optimization (DTCO), the process of co-optimizing design and technology to achieve the best tradeoff between manufacturing cost and design efficiency, is being proposed to scale effectively below the 32 nm node [21] . DTCO has been very successfully applied to SRAM bitcell scaling for a few decades, and researchers are proposing to extend the DTCO process to include standard cell and analog components as well, resulting in a wholesome process technology definition [25].

During a DTCO process, process engineers communicate process capabilities and limitations codified in the form of RDRs and allowable pattern constructs [23][24] to the designers. A pattern construct is a rectangular piece of layout surrounding a target feature of interest [26]. The size of construct's rectangular region is determined by the lithographic region of influence (Figure 2.14). With initial RDRs and constructs, designers attempt to build a few fundamental leaf cells, bitcells and standard cells, in an efficient manner, following which they give feedback to process engineers on what technology features were helpful and what more is required to make the cells more efficient. The goal of such a DTCO process is to identify the smallest set of unique constructs that is most useful for designers, as minimizing the number of unique constructs significantly improves manufacturability [27]. Iterating in this manner, DTCO enables process engineers and designers to best tradeoff manufacturing cost and design efficiency. In the next chapter we undertake the DTCO process for a state-of-the-art 14 nm technology.

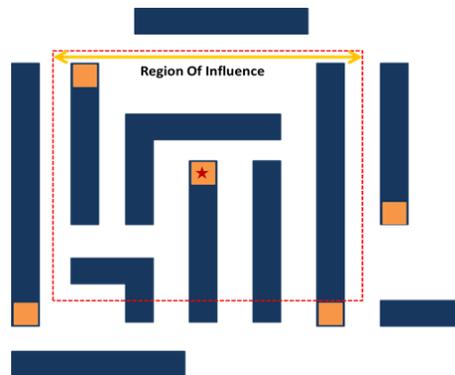

**Figure 2.14 A pattern construct is defined by its target feature (e.g. V1), a rectangular region of influence (e.g. 5 pitch wide square) and the layers holding its relevant polygons (eg. V1 and M1) [24][26].**

## 2.3 Summary

In this chapter we reviewed several challenges facing affordable CMOS scaling beyond the 20 nm node, in the realms of design, technology and CAD. To offset soaring manufacturing cost it is critical to scale SoCs efficiently as required by Moore's law. However, conventional design scaling techniques are ineffective due to the introduction of several new technology elements and increasingly restrictive patterning. As the premise for affordable scaling is being challenged, the 14 nm CMOS technology node emerges as an inflection point. While design technology co-optimization at the leaf cell level has been reasonably successful in making the 20 nm node cost-effective, its ability to make an extremely patterning restricted 14 nm node profitable is yet to be explored.

# 3 Leaf Cell Design Technology Co-Optimization for Sub-20 nm CMOS

Design technology co-optimization (DTCO) has proven to be an effective technique to balance manufacturing cost and design efficiency at the 20 nm technology node. The DTCO process is typically driven by designers in conjunction with process engineers to converge on an efficient bitcell and standard cell, the fundamental SoC building blocks. As the CMOS process stack is not expected to change much until the 7 nm node, we undertake the DTCO process in a foundry 14 nm CMOS process with artificially imposed 10 nm and 7 nm technology assumptions. As discussed briefly in Section 2.1.1, lithographic patterning is the most daunting technology scaling challenge and we explore a few promising next generation lithography (NGL) options next.

## 3.1 Patterning Choices at Sub-20 nm CMOS

Different patterning options, their challenges and readiness for high volume manufacturing (HVM) is summarized in Table 3.1. Multi-patterned 193 immersion (193i) lithography is set to drive high volume manufacturing in the 14 nm node and is the most mature option to drive the 10 nm node [28]. Extreme Ultra Violet (EUV) lithography (EUVL) is still battling technical and economic challenges and is not expected to be used until the 7 nm node, by which time EUVL might need to be double patterned [29]. Given the high complexity and cost associated with these techniques, lithographers are also actively exploring a few alternative next generation lithography (NGL) options, such as self-aligned multiple patterning [30] and Directed Self-Assembly (DSA) [31]. Realistically, 10 nm and 7 nm processes are expected to use different lithography options depending on the design patterns seen on each layer. Such design patterns can be broadly categorized as pure gratings, structured 1D gratings and compound 2D gratings, in increasing order of patterning complexity (Figure 3.1). While there have been proposals to build designs exclusively out of pure grating patterns [32], our experiments reveal that pure grating-based cells are area and power inefficient (Figure 3.2). Furthermore, these cells i) have minimum M1 area input pins, that raises serious process concerns and ii) minimum width (1X) power rails that are susceptible to increased IR drop and electro-migration. Running multiple straps of wider metals for power and ground over the 1X metals could help alleviate the IR drop problem but that would significantly decrease the number of tracks available for signal routing. Hence, it is essential to take an all-inclusive view to address this challenge, for which we undertake a layer-wise DTCO.

Table 3.1 Sub-20 nm node patterning techniques, their challenges and readiness for high volume manufacturing (HVM).

| Technique | Challenges | Ready for HVM |
|---|---|---|
| EUV (13.5 nm) | Very expensive, Mask defects, Low Wph | 7 nm |
| LELELE (193i) | Expensive, Overlay error | 14 nm, 10 nm |
| SADP/SAQP (193i) | Moderately Expensive, Restricted-2D, 1D layout | 14 nm → 7 nm |
| DSA (193i) | Cheap, 1D layout | 10 nm → 7 nm |

Considering different patterning choices and associated restrictions, we assume that poly, fin, diffusion and diffusion contacts (CA) are pure grating layers. Such an assumption is made to minimize process variations as these layers form the FinFET device. Restricting the gate contact (CB) to be a grating would severely limit design efficiency and hence is restricted to be a 1D layer. The directionality and restrictiveness of back end of line layers (BEOL) can be used to tradeoff between design efficiency and manufacturing cost. Of the BEOL layers, 1X layers are the most critical, particularly M1. During leaf cell DTCO we assume M1 can be a structured 1D grating or a compound 2D grating. M2 and M3 layers are allowed to be structured 1D gratings to reduce manufacturing cost, as summarized in Table 3.2. Furthermore, during DTCO, we attempt to minimize the number of unique constructs, with the goal of improving manufacturability.

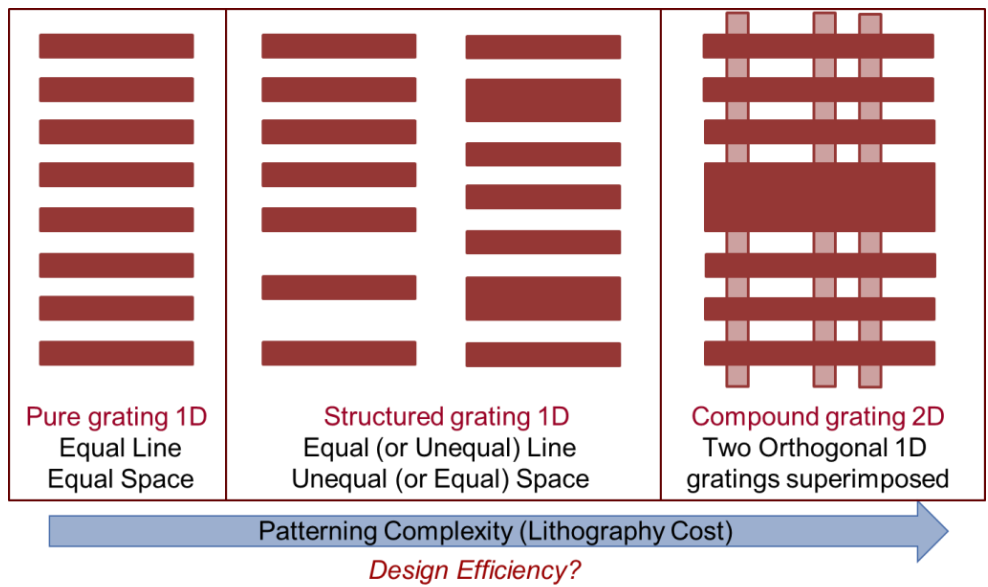

**Figure 3.1 Patterning choices at 14 nm and beyond.**

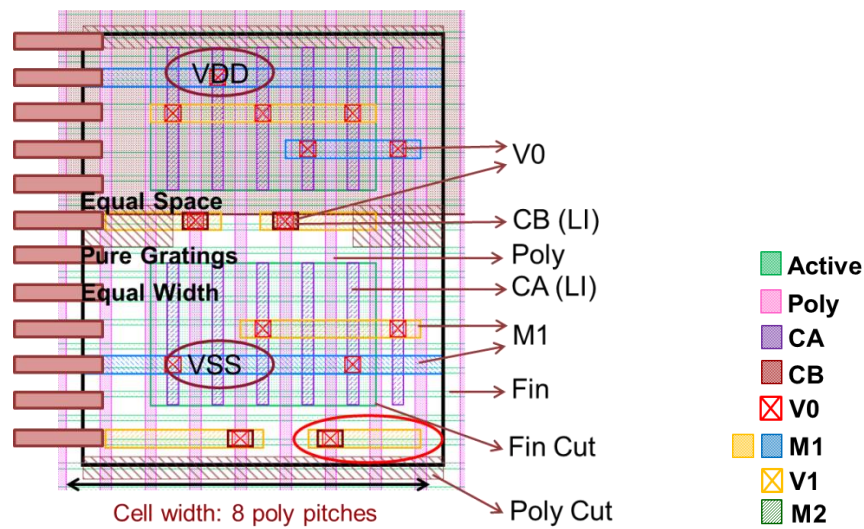

**Figure 3.2 Pure grating cells are inefficient for design and process integration.**

Table 3.2 Patterning restrictions assumed for Sub-20 nm DTCO.

| Layer | Patterning Restrictions |
|---|---|
| Poly, Active, CA | Pure grating 1D |
| CB | 1D |
| M1 | Structured 1D or Compound 2D |
| M2, M3 | Structured 1D |
| Vias | Self aligned |

## 3.2 Sub-20 nm Bitcell Design

Bitcells form the most fundamental building block of an SRAM and historically drive the DTCO process. Bitcell DTCO is geared to converge on a technology definition that enables the design of a compact, high performance, stable, low $V_{min}$ and high yielding bitcell. 6 Transistor (6T) bitcell has been used widely for nearly two decades, but recently have required read and write assist circuits for improved stability that makes them less attractive. Chang et al., proposed the 8 transistor (8T) bitcell and demonstrated that at the array level 8T bitcell provides a more high performance, area and power efficient alternative [33]. Therefore, we subject the scalable 8T bitcell to DTCO.

All sub-20 nm processes feature FinFETs that require transistors to be sized discretely, a marked departure from planar FETs. During DTCO we converged on the 8T bitcell circuit shown in Figure 3.3. The requirement to avoid notches in the RX layer dictates that both devices that share diffusion need to have the same number of fins. During layout of the bitcell we learned that running M1 vertical (M1V) and M2 horizontal (M2H) did not result in an area efficient bitcell. Furthermore, running BEOL layers in a bidirectional manner is typically discouraged for bitcells due to yield concerns. The two area efficient layout configurations we converged on after DTCO were M1HM2V and M1HM2HM3V, as shown in Figure 3.4. The two bitcells were laid out with the restriction that bitlines should be kept on M1 for high performance. The two bitcells were 16 M1 tracks tall and 2 poly pitches wide. Details of other bitcells that were explored during DTCO are presented in [34].

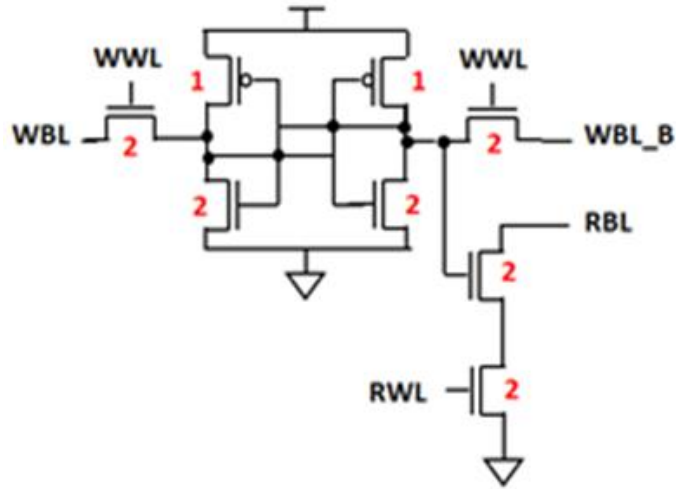

**Figure 3.3 8T-Bitcell circuit.**

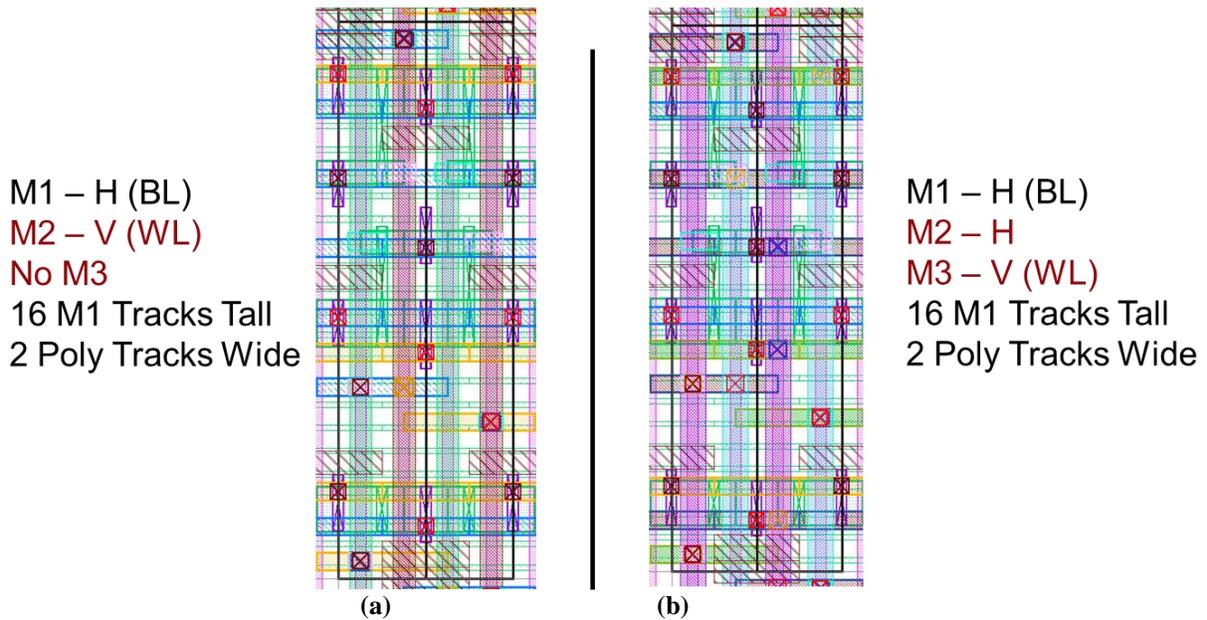

M1 – H (BL)
M2 – V (WL)
No M3
16 M1 Tracks Tall
2 Poly Tracks Wide

M1 – H (BL)
M2 – H
M3 – V (WL)
16 M1 Tracks Tall
2 Poly Tracks Wide

(a)    (b)

**Figure 3.4 8T-Bitcell layout topologies (shown as 1x2 bitcell array): (a) M1HM2V, (b) M1HM2HM3V.**

## 3.3 Bitcell Array Boundary and its Implications on Standard Cell Design

As conventional geometric shrink of standard cells is not possible below the 32 nm CMOS node, it is essential to DTCO standard cell logic. However, prior to starting the standard cell DTCO, it is important to understand any interdependencies that may exist between standard cells and bitcells, especially at the pattern boundary between them (Figure 3.5). Owing to severe patterning restrictions it is very difficult, if not impractical, to support significantly different process options for logic and bitcells. Extremely restricted FinFET device layers such as fin and poly are forced to be identical for bitcell and logic; i.e., they follow the same pitch and track plan. In order to understand which other layers might have to follow the bitcell track plan, we investigate the bitcell array boundary in detail. Interestingly, bitcell arrays require a few poly pitches to strap power/ground and for placing pins for write and read bitlines (Figure 3.6). This space in the bitcell array boundary could be effectively used to transition to a different track plan for other layers such as diffusion, CB, M1, M2 etc. Next, we undertake standard cell DTCO following the bitcell's fin and poly track plan along with patterning restrictions mentioned in Table 3.2.

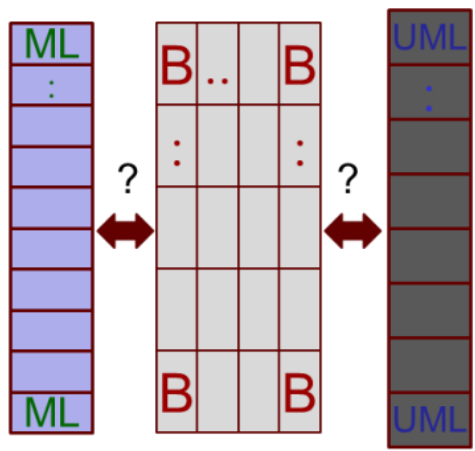

**Figure 3.5 Bitcell-Standard cell logic boundary. B is bitcell, ML is matched logic and UML is unmatched logic.**

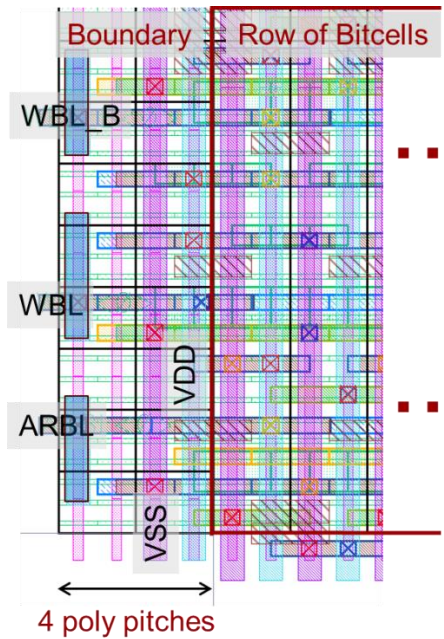

**Figure 3.6 Bitcell array boundary.**

## 3.4 Sub-20 nm Standard Cell Design

Standard cell (or logic cell) libraries are used to implement the digital functions of a chip. As standard cells are not typically DTCO'ed, we formulate the following goals for standard cell DTCO process.

•       Minimizing cell area is driven by technology scaling requirements and is an important design goal. A crucial standard cell parameter that determines the cell area is the "cell height," also called "track height" (Figure 3.7). Track height is fixed for all cells in a given standard cell library, and our goal is to minimize track height without compromising other goals listed below.

•       Maximize active area efficiency, which is defined as the total active area in the cell divided by cell area. In a FinFET process, active area efficiency can be expressed as fin efficiency:

Fin Efficiency = (# of fins forming active devices in a logic cell) / #total of fins.

•       Minimize contacts, vias and Metal2 (M2) in the standard cell.

•       All layers follow restrictions mentioned in Table 3.2. Additionally poly and fin layers follow the same track plan as the bitcell.

•       Wider power rail structures for electromigration tolerance.

•       Minimize parasitic capacitance.

- Scalable circuit and layout.

Satisfying these goals, we create a compound grating-M1 based standard cell (10T_BiDir) and structured grating-M1 based standard cell (10T_UniDir).

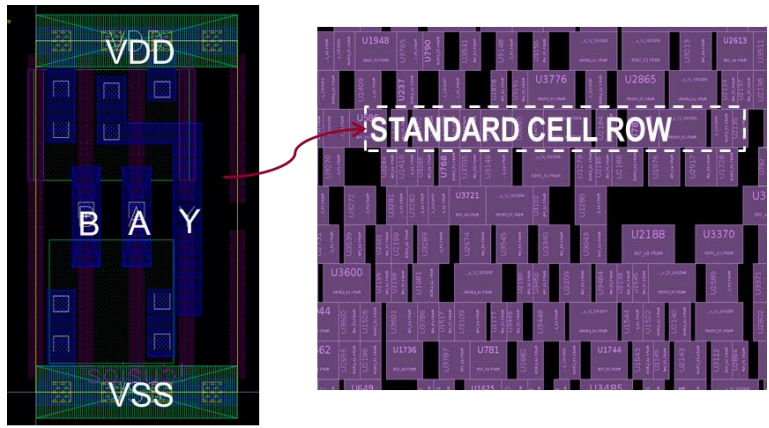

**Figure 3.7 Standard cell assembled into a block.**

### 3.4.1 Compound Grating Metal1 (M1) based Standard Cell

A compound grating-based bidirectional standard cell layout, called bidirectional cell layout for simplicity, contains M1 shapes patterned using two orthogonal grating patterns, as shown in the 10T_BiDir NAND2_X1 cell in Figure 3.8. In this 10T_BiDir standard cell layout we have attempted to retain most of the attributes of a typical bidirectional standard cell layout, such as power rails shared across the adjacent rows of standard cells and input pins pushed towards the center of the cell. However, to comply with patterning restrictions in sub-20 nm processes while still retaining its design efficiency, the bidirectional standard cell layout has had to evolve further. Its key attributes are as follows:

- 10T_BiDir contains a robust power rail structure made of local interconnects and metals. The use of this novel and robust power rail structure, instead of the traditional M1 taps, serves as a key feature of 10T_BiDir.

- CA taps are used to connect the source/drain of the transistors to the CB power rail.

- CB is used to contact the poly to make input pin connections. Therefore, CB need not be very dense and does not push lithography resolution limits. We leverage this and do not restrict

CB to be a grating, which in turn helps us greatly improve the design efficiency of our standard cells by at least 10%.

- This cell is also 10 tracks tall and has a fin efficiency of 66.67%.

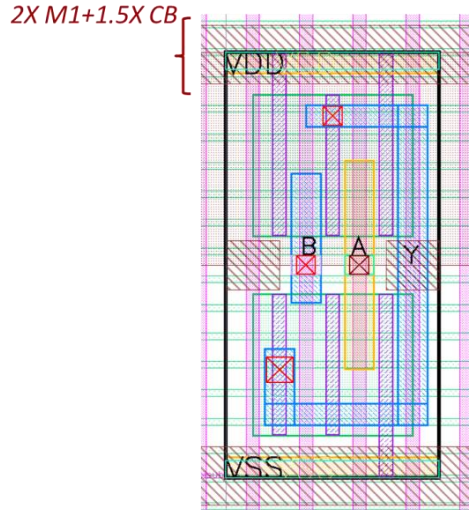

**Figure 3.8 2D-M1 based Standard Cell – 10T_BiDir (NAND2_X1).**

### 3.4.2 Structured Grating Metal1 (M1) based Standard Cell

Traditionally, unidirectional-M1 standard cells had been created by replacing bidirectional M1 with a combination of vertical M1, via1 (V1) and horizontal M2 [37]. Such an approach was pursued, as the process options in previous technology nodes were not amenable to creating an efficient unidirectional standard cell. The key attributes of the structured-grating based unidirectional-M1 standard cell (10T_UniDir) layout shown in Figure 3.9 are listed below:

- M1 in the cell is perpendicular to poly and M2 is parallel to poly. The choice of these orientations comes from the fact that M1 has to be perpendicular to the first extensively used local interconnect level (CA is parallel to poly) to maximize and ease connectivity.

- To minimize M2 usage in the cell, the CA layer is used – beyond its envisioned usage – to connect N type and P type transistors, making it robust with less parasitics.

- The input pins connect to poly at either the center or the bottom edge of the cell layout. Pin locations have been strategically chosen to enable the use of CA to connect N type/P type transistors, as the presence of all gate contacts in the center of the cell will disallow any CA connection between N type/P type transistors. Furthermore, this allows the input pins to be spaced further away from each other, improving the pin access.

- The inputs pins are wider to avoid process risk prone minimum metal area shapes.

- The cell contains 2X wide power rails to improve electromigration tolerance.

- CB is used to contact the poly to make input pin connections. Therefore, CB need not be very dense and does not push lithography resolution limits. We leverage this and don't restrict CB to be a grating, which in turn helps us greatly improve the design efficiency of our standard cells by at least 10%.

- The unidirectional cell tiles without mirroring, unlike a conventional standard cell which requires mirroring the cell to tile.

Meeting these design/process constraints and 14 nm design rules, the structured grating-based standard cell is 10 tracks tall and has an active area efficiency of 66.67%.

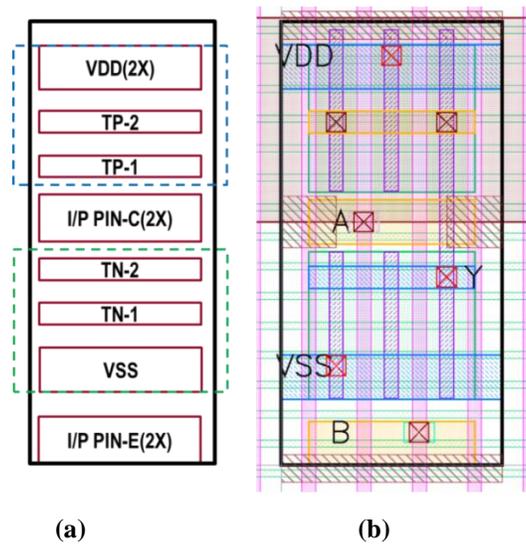

(a) (b)
**Figure 3.9 10T_UniDir Standard Cell: (a) Track Plan, (b) Layout.**

### 3.4.3 Sub-20 nm Standard Cell Library Comparison

We analyzed the proposed standard cells, namely, compound grating (10T_BiDir) and structured grating (10T_UniDir, 9T_UniDir) for design efficiency and manufacturability. We created a 40 cell representative standard cell library in the IBM 14 nm process. To study their manufacturability, we assembled two random logic blocks each made of cells from 10T_UniDir and 10T_BiDir libraries, respectively. SEMs from testchips fabricated at IBM show good patterning fidelity for both 10T_BiDir and 10T_UniDir (Figure 3.10). Furthermore, analysis of the number of unique constructs and process variability indices (PVI) quantify the improved manufacturability of these two standard cell architectures [37].

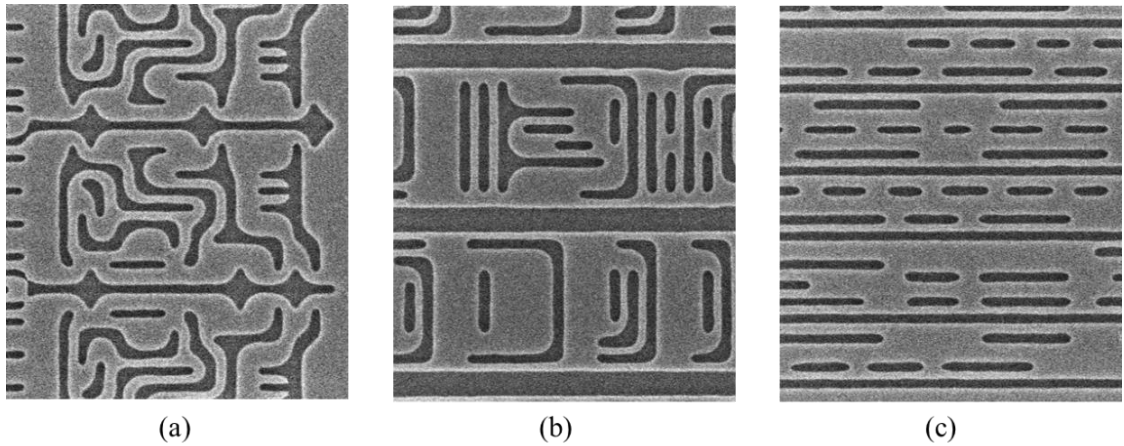

**Figure 3.10 M1-SEMs from different standard cell layouts. (a) 32 nm like layout style, (b) 10T_BiDir random logic block, (c) 10T_UniDir random logic block.**

The cells in 10T_UniDir and 10T_BiDir were also characterized to create timing libraries (.lib) and physical abstracts (.lef) for building ASIC blocks. The two standard cell libraries were analyzed for metrics at the library level and block level. Simulations at the library level indicate that both 10T_UniDir and 10T_BiDir exhibit similar area, power and performance (Figure 3.11). We use the BEOL stack in Table 3.3, with unidirectional M2 and M3 and restricted-2D M4 and M5, during block-level analysis. Unidirectional metal stack improves manufacturability by limiting the number of layout patterns while also allowing the use of cost-effective patterning techniques, such as, DSA and SADP. Restricted BEOL stack also allows us to use simplified gridded auto-routers, improving design quality and design turnaround time. However, unidirectional metal stack inhibits conventional via redundancy techniques requiring us to use alternative via redundancy schemes such as local loops [36]. Physical synthesis results from a 32 bit multiplier indicate that the two libraries have similar design efficiency [37]. During physical synthesis we also observed that cells in 10T_UniDir have better pin access compared to 10T_BiDir as shown in Figure 3.12 [38]. The comparison between 10T_UniDir and 10T_BiDir is summarized in Table 3.4.

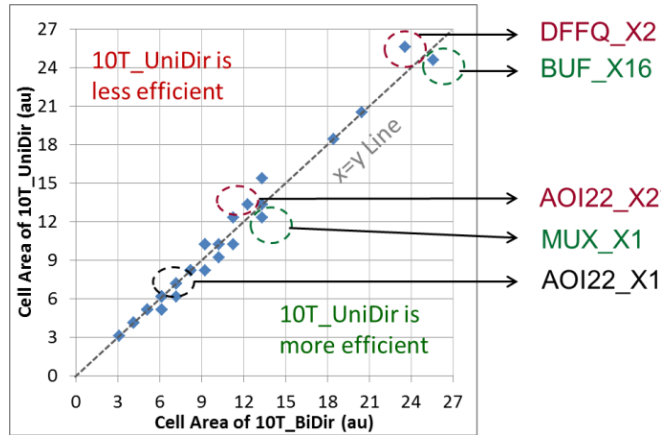

Figure 3.11 Design efficiency comparison of 10T_UniDir and 10T_BiDir standard cell library.

Table 3.3 BEOL Stack for 10T_BiDir and 10T_UniDir.

| Layer | 10T_BiDir BEOL Stack (Direction/Pitch) | 10T_UniDir BEOL Stack (Direction/Pitch) |
|---|---|---|
| PC | V/y | V/y |
| M1 | H/x & V/y | H/x |
| M2 | H/x | V/y |
| M3 | V/x | H/x |
| M4 | H/1.5x and V/3x | V/1.5x and H/3x |
| M5 | V/1.5x and H/3x | H/1.5x and V/3x |

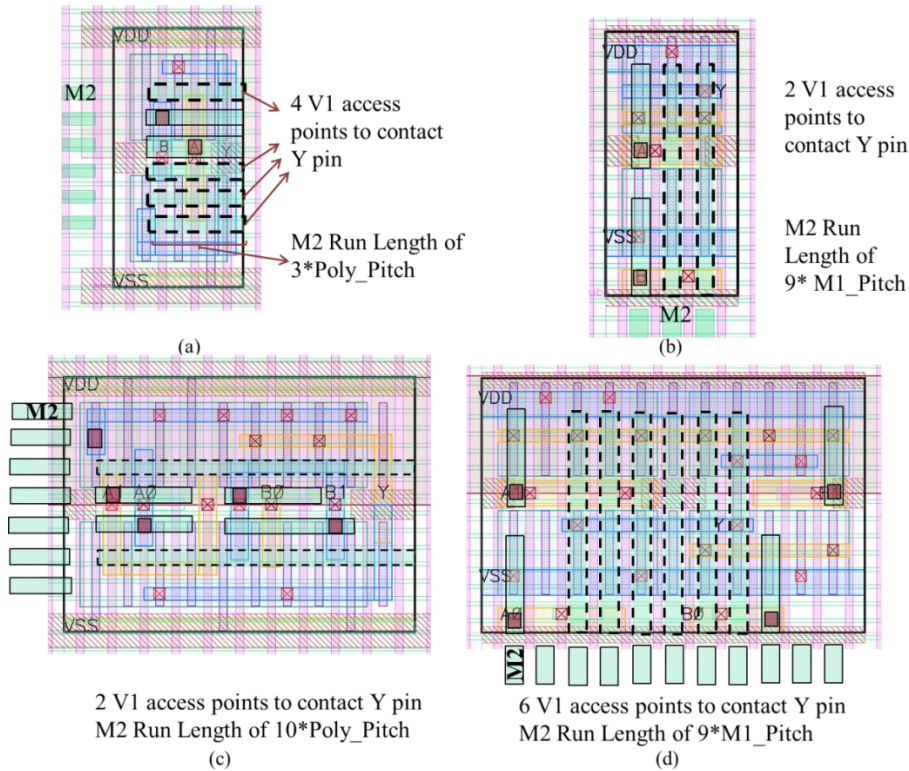

Figure 3.12 Pin access for 10T_BiDir and 10T_UniDir cells. (a) NAND2_X1 in 10T_BiDir ; (b) NAND2_X1 in 10T_UniDir ; (c) AOI22_X2 in 10T_BiDir ; (d) AOI22_X2 in 10T_UniDir.

**Table 3.4 Design efficiency comparison table of 10T_UniDir and 10T_BiDir.**

| Attributes | 10T_UniDir | 10T_BiDir |
|---|---|---|
| Cell Height (M1 tracks) | 10 | 10 |
| Active area Efficiency | 66.66% | 66.66% |
| Power Rail Efficiency | 2/10 | 2/10 |
| NAND2_X1 Transistor Eff. | 2/4 | 2/4 |
| DFFQ_X1 Transistor Eff. | 13/25 | 13/23 |
| AOI22_X1 Transistor Eff. | 4/6 or 4/7 | 4/6 or 4/7 |
| INV_X1 Transistor Eff. | 1/3 | 1/3 |
| Redundancy Possibility(1-5) | 3 | 4 |
| Block Level Routability(1-5) | 4 | 2 |

1 is bad
5 is good

Active area Efficiency $\cong$ Active Area / Total Cell Area,
Transistor-Efficiency = [(# active poly-gates) X poly_pitch] / Cell width,
Power Rail Efficiency = # Power Rail Tracks / Total # of tracks in standard cell.
Redundancy and Block Level Routability are ranked from 1 to 5 with 5 being the best and 1 the worst.

## 3.5 Leaf Cell DTCO Inadequate to Drive Affordable Scaling Below 20 nm CMOS

Given our patterning assumptions, our leaf cell DTCO has converged on the best possible bitcell and standard cell. While comparing the area efficiency of our DTCO'ed 14 nm standard cells to a similar cell in the 32 nm node, we observed an area scaling of ~60% (as compared to the expected ideal Moore's area scaling of 75%). Similarly, for the SRAM bitcell, we scaled by about 61% compared to a similarly sized bitcell in the 32 nm node. Area scaling achieved by leaf cell DTCO, while being better than the anticipated 40%-50% scaling from 32 nm to 14 nm [12], still is inadequate to meet ideal node to node area scaling of 75%. This is a huge concern as bitcell and standard cell scaling has traditionally driven SoC scaling. Especially, as SRAMs form up to 80% of a modern SoC area, it is important to understand the impact of poor bitcell area scaling on overall SRAM scaling.

### 3.5.1 Sub-Optimal Compiled SRAM Block Scaling

Modern SoCs, unlike microprocessors, use tens to hundreds of different SRAM blocks. To satisfy these diverse SRAM needs, SoC designers use SRAM compilers. SRAM compilers work by tiling hard IP (SRAM periphery) leaf cells. The efficiency and diversity of the leaf cells single handedly determines the quality of SRAMs that are generated from SRAM compilers. While the compilation approach has served designers well over the past few decades, it is increasingly proving to be inefficient as we scale to nodes below 32 nm. Restrictive patterning and associated process variations have rendered several leaf cell topologies, such as post-charge row decoders, inefficient or unusable [39]. The cost of characterizing leaf cells in silicon is proving to be expensive and impractical.

Furthermore, limited by hard IP building blocks and associated custom design and pre-characterization, a commercial SRAM compiler can only create discrete memory sizes with a pre-determined SRAM architecture [40]. This compilation strategy not only limits the possibility of application-specific customization but also hinders comprehensive design space exploration, leading to a sub-optimal IP. For instance, it is not uncommon to see a 256*8 (256 word SRAM with 8 bits per word) and a 256*64 SRAM have the same memory periphery, specifically row decoders. Using row decoders sized for 256*64 SRAM in a 256*8 SRAM results in sub-optimal power and area.

To observe the impact of patterning restrictions on SRAM compilers we built a compiled 1R-1W 1KB SRAM. We used 8T-bitcells and static CMOS based periphery for improved manufacturing robustness. The resulting SRAM block, shown in Figure 3.13, was observed to be robust but inefficient (described in detail in Section 5.5.1). This exercise illustrates the need to look beyond co-optimizing layout and process technology to improve design efficiency of embedded memory blocks.

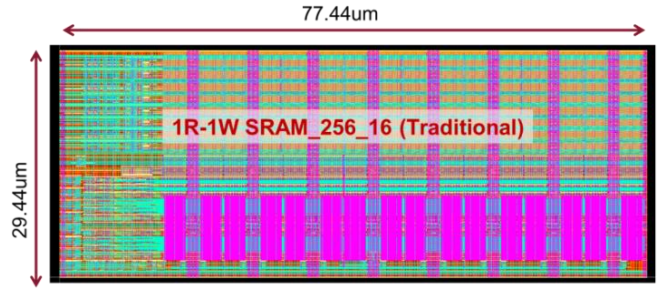

**Figure 3.13 Top-level layout and floorplan of a compiled two-port SRAM.**

## 3.6 Summary

Encouraged by the success of design technology co-optimization (DTCO) in the 20 nm node, we undertook 14 nm DTCO in collaboration with IBM. Two fundamental leaf cells, SRAM bitcell and standard cell, were used during DTCO. Working on a 14 nm process with several new technology elements, we converged on a manufacturable and efficient 16 track, 8 transistor SRAM bitcell and 10 track standard cell logic (10T_UniDir and 10T_BiDir). Leaf cell DTCO improved manufacturability while achieving a node to node area scaling of ~60%, compared to a similar leaf cell in the 32 nm node. With critical pitches scaling only by a factor of 0.8 (Section 2.1.1), leaf cell DTCO almost achieves the best area scaling possible by working at the circuit and layout level. However, the limited scope of DTCO to co-optimize circuits, layout and process technology proves to be inadequate to meet ideal node to node area scaling requirements (of 75% from 32 nm to 14 nm) for the heavily patterning restricted 14 nm node.

# 4 Broadening DTCO to Exploit the Challenges of Sub-20 nm CMOS

From the previous chapter it is clear that leaf cell DTCO is not sufficient to drive affordable CMOS scaling below the 20 nm node. To best exploit the patterning impediments it is imperative to broaden the DTCO process to include micro-architecture and CAD methodologies along with circuit, layout and process technology. As embedded memories, specifically SRAMs, constitute more than 50%-80% of a modern SoC area [12], it is essential to pursue a holistic DTCO process for SRAM blocks for affordable CMOS scaling. A more holistic co-optimization process that is aware of the application that uses the SRAM, and also how SRAM blocks are generated for integration into an SoC, can enable the generation of more efficient and manufacturable SRAMs in a productive manner (Figure 4.1).

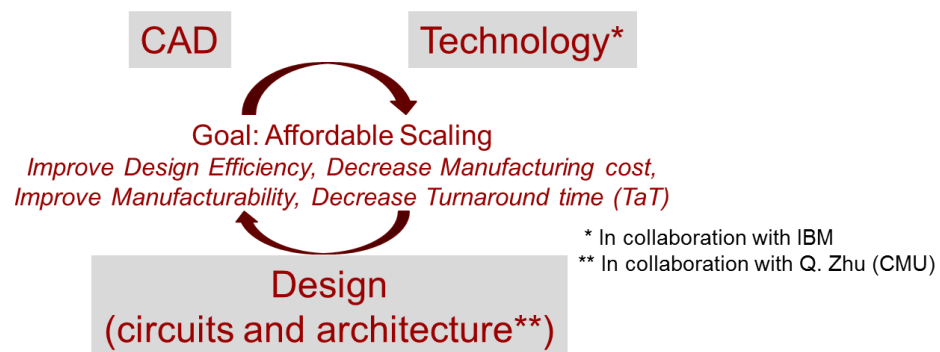

**Figure 4.1 Holistic design technology co-optimization (DTCO) includes micro-architecture and CAD along with circuits, layout and process technology.**

## 4.1 Holistic DTCO Essential to Improve SRAM Efficiency

The partitioning of logic and memory with the conventional Von Neumann computing paradigm is based on storage-only "dumb" embedded memories that are becoming increasingly less efficient as the *on-chip* data transport energy now exceeds the energy consumed by logic computation [41]. An effective solution that addresses this challenge, while still retaining a simplified memory abstraction for the system, is to enhance the embedded memory functionality by integrating application-specific computation into the memory abstraction. Embedded memory IP startup Memoir Systems [42] offers such a customized embedded memory, referred to as algorithmic memory, to alleviate the performance gap between the processor and embedded memories. Algorithmic memory uses existing single port SRAM blocks, generated from SRAM compilers, along with specialized logic to create multi-ported SRAMs, thereby increasing the memory operations per second (MOPS). This solution, however, still relies on the ability of SRAM compilers to generate efficient single-port SRAMs, which as noted in the previous chapter is very difficult, if not impossible in deeply scaled technology nodes.

Several designers and researchers [43][44][45] have demonstrated the significant energy and area benefits of co-optimizing SRAM circuits with application design, by making application-specific customizations to SRAM blocks. However, these customizations are not used widely due to their high design cost and effort, owing to their specialized and design intensive nature. Application-specific SRAMs/Register-Files (RFs) are not integrated into mainstream chip design methodologies since they cannot be robustly generated and validated using SRAM/RF (referred to as SRAM in remainder of this dissertation) compilers. To expand the utility of such efficient application-specific SRAMs to a wide variety of products it is necessary to create efficient CAD methodologies that can drive holistic DTCO of SRAMs.

## 4.2 Smart SRAM Synthesis Enables Holistic DTCO

Extreme patterning restrictions significantly limit the diversity of efficient leaf cells that a designer feeds into a traditional SRAM compiler, rendering the approach inefficient. Synthesis, unlike compilation, can enable holistic DTCO by automatically generating an application-specific SRAM block that is built from robust circuit primitives and customized to user specifications, thereby, guaranteeing manufacturability and efficiency. Customized to application needs and enabled by technology, such "smart memory" enables fine grained application-specific optimization to the embedded memory circuits, that is otherwise impractical with the traditional SRAM and standard cell hard-IP based design flows (Figure 4.2). Furthermore, these smart memories can be described via an RTL interface and implemented using physical synthesis flows, thereby dramatically improving design productivity.

An example smart memory block is shown in Figure 4.2 with three attributes. First, the smart memory block retains the traditional memory interface, an interface system architects prefer. The convenient memory interface enables smart memories to be easily integrated into any system's architecture. Second, smart memories are synthesized using two building blocks, augmented bitcell arrays (BA+), that represent our embedded memory building block, and efficient standard cells for peripheral and application-specific logic. An augmented bitcell array (BA+) contains a small array of bitcells with very basic periphery to make it compatible to physical synthesis (more in Section 4.4). Third, smart memory blocks can be synthesized automatically from RTL to GDS using commercially available high productivity physical synthesis flows. While the benefits of smart memory synthesis are intuitive, their robust manufacturability is not. Hence, we study the manufacturing feasibility of smart memories in the next sub-section.

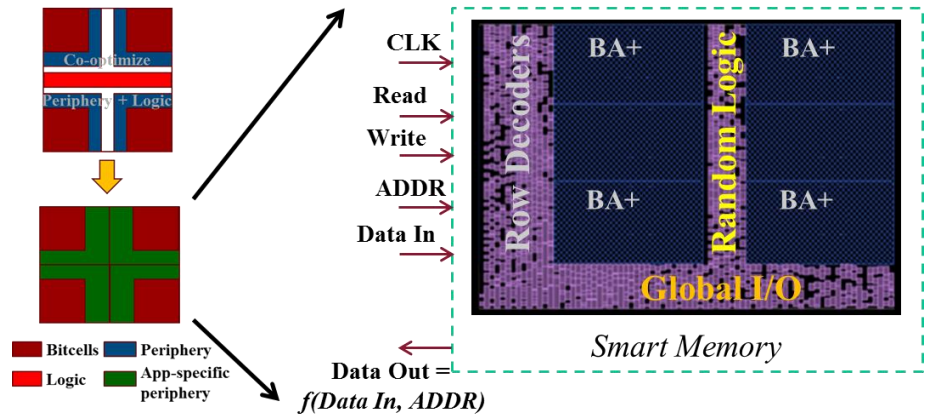

**Figure 4.2 Smart memory.**

## 4.3 Synthesis Exploits Opportunities in Restrictive Patterning

The most area efficient smart SRAM designs will require placement of standard cell logic and SRAM bitcells within close proximity (2-4 poly pitches). For a modern foundry process, such as 32 nm, this is infeasible since the logic and bitcells are lithographically incompatible. In such situations, the logic and bitcells are separated by 8-12 poly pitches from each other to ensure reliable patterning. Therefore, any standard cell logic added within a memory block will incur an area overhead for this patterning separation. However, as we scale to nodes below 20 nm, extreme patterning restrictions require bitcell and standard cell logic to be designed from the same set of restricted design rules and pattern constructs. This suggests that for 14 nm and subsequent nodes we can construct logic and bitcell patterns that are lithographically compatible, without incurring an area overhead.

To validate this hypothesis we placed a design-efficient, manufacturable and extremely regular standard cell [37] in close proximity with optimized embedded memory bitcell arrays on a 14 nm patterning test-chip and compared its printability with conventional (non-lithographically compatible) standard cell logic placed near the same bitcell array. We then observed the impact of neighborhood on bitcell patterning. Since all FEOL layers are very restricted and grating-like, we inspected Scanning Electron Microscope (SEM) images from the Metal1 (M1) layer. Figure 4.3 illustrates that conventional standard cell layout patterns disrupt SRAM bitcell patterning, whereas placing extremely regular standard cells in close proximity (2-4 poly pitches) does not impact bitcell patterning. This experiment demonstrates, on silicon, the fine grained integration of computation logic in embedded memory for a state-of-the-art 14 nm process technology. Furthermore, as the standard cell logic and bitcell arrays were designed from a small set of unique pattern constructs, it significantly improves manufacturability of smart memories.

This new technology capability allows us to synthesize efficient smart memories with tightly coupled logic and embedded memory. Such smart memory synthesis would require new circuit building blocks and customized physical implementation tools and flows for the rapid creation of efficient designs by allowing for the control of embedded memory architecture parameters, such as type of bitcell, number of bitcells per bitline, etc. We describe the smart memory building blocks in the next sub-section.

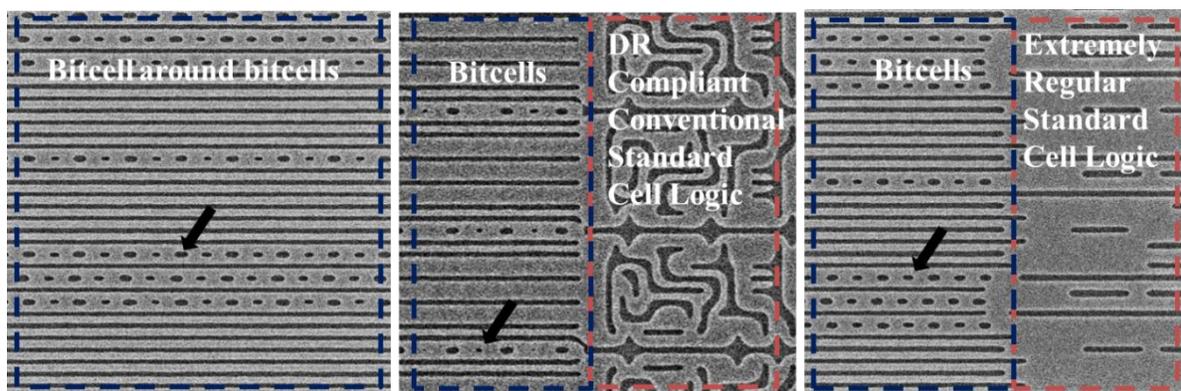

**Figure 4.3 Metal-1 SEMs from a 14 nm patterning testsite showing lithographic compatibility between array of bitcells (BA+) and extremely regular standard cell logic.**

## 4.4 Robust Building Blocks for Smart SRAM Synthesis

Our synthesized smart memory block is broadly comprised of two building blocks, an augmented bitcell array (BA+) to store and standard cell logic to compute. The smart memory function is specified in RTL and is automatically constructed using custom modifications to physical synthesis tools and flows. Physical synthesis of a smart memory block requires its building units, namely BA+ and standard cells, to have a robust timing abstraction (.lib). Creating a .lib for standard cells with static output nodes is straight forward, whereas, capturing the behavior of bitcell arrays with dynamic output nodes is a challenge. This challenge is tackled by designing simple static wrappers around bitcell arrays, which make their input/output interface static, thereby making them compatible with physical synthesis tools. The 8-transistor (8T) bitcell based 16x16 BA+ is shown in Figure 4.4 as an illustration. It contains a small size bitcell array (e.g. 16 entries with 16 bitcells each), local sense for reading, and clock enabled wordline driver buffers without any decode functionality. The local sense merges the internal read bitlines and feeds it to a tri-state global bitline driver. This tri-state output driver ensures that multiple BA+ cells can still share an array read bitline (ARBL), while avoiding the traditional pre-charged dynamic ARBL structure. Such a BA+ design allows itself to be integrated with any kind of customized (or non-customized) global peripherals while making it compatible with physical synthesis.

Different BA+ designs, subsequently, can be automatically generated to provide a range of synthesis options, such as different number of bitcells in a wordline (16x8, 16x16) or bitline (32x16, 64x16), different bitcell sizing (pull down to access ratio), and different bitcell layout configurations (pushed layout, conservative layout etc.). A few BA+ cells were designed and characterized in IBM 14 nm process. Figure 4.5 illustrates the performance, area and energy*delay for different "BxW BA+" that have different number of bitcells in a bitline (B) and wordline (W). BA+ of size 32x16 and 64x8 store the same number of bits (0.5Kb), but exhibit different design tradeoffs. 32x16 has lower latency and consumes lesser area, but 64x8 is more energy efficient. This makes the correct choice of BA+ critical to meet design targets. Given a bitcell circuit and its layout, such as, 6T, 3T-1D eDRAM, 9T/10T TCAM, a library of BA+ could potentially be automatically generated by an in-house BA+ generator. From such a comprehensive library of BA+, an efficient smart memory implementation meeting design specifications can be automatically explored and synthesized using the proposed smart memory synthesis framework (SMSF).

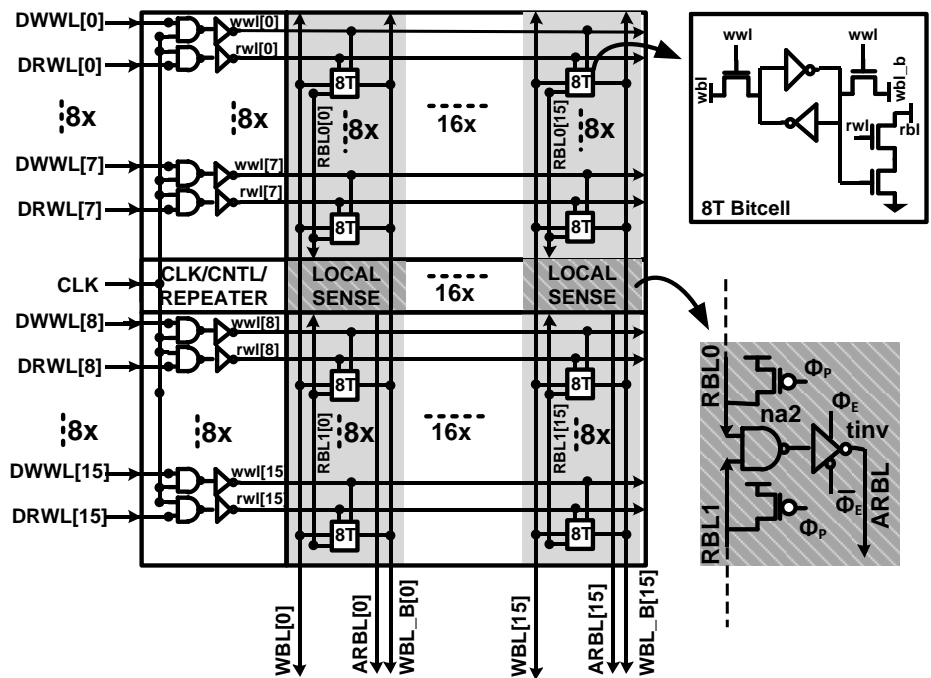

**Figure 4.4 An augmented 16x16 Bitcell Array (BA+) circuit with wordline driver, local sense, clock and control logic.**

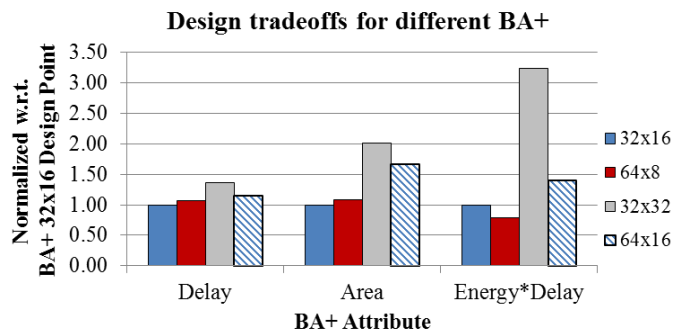

**Figure 4.5 BA+ design space illustration (graph normalized w.r.t. BA+ 32x16 design point for ease of understanding).**

## 4.5 Smart Memory Synthesis Framework (SMSF)

Our generic SMSF generates optimal physical implementation of an application given a variety of architecture and physical parameters, as illustrated in Figure 4.6. It consists of an SMSF frontend that is used for design space exploration and synthesizable RTL generation and an SMSF backend that maps the RTL to an efficient physical implementation targeted to meet design specifications. Generic SMSF is customized for each application to generate the most efficient designs.

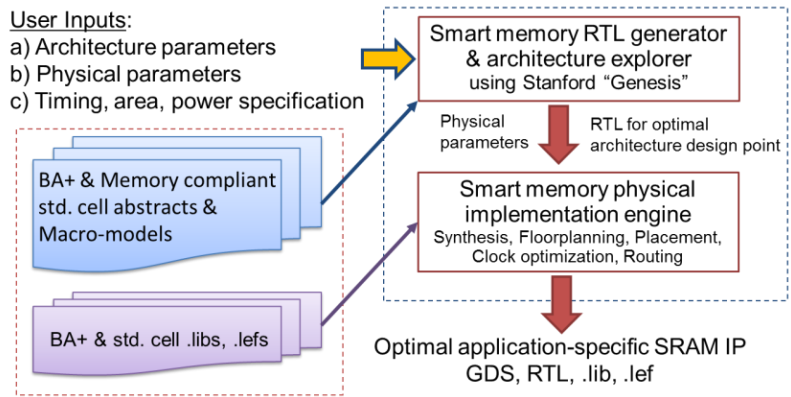

**Figure 4.6 Generic smart memory synthesis framework (SMSF).**

### 4.5.1 SMSF Frontend

Our generic SMSF frontend is based on Stanford's Genesis framework. Genesis allows RTL designers to encapsulate all parameters of a specific application's hardware into a single Verilog/Perl (.vp) template file. For user prescribed design parameters, the Genesis tool can then automatically generate a synthesizable hardware description in Verilog [46]. Furthermore, we have augmented the Genesis framework to automatically explore the design space in search of the candidate design points that best meet the desired user specification. To accomplish this task the SMSF frontend requires the following inputs: a) design-specific micro-architecture parameters like embedded memory size, fixed-point precision, number of embedded memory partitions, number of parallel computation units etc., b) physical parameters like floorplan-aspect-ratio (height/width), c) design power-performance-area specifications and d) power-performance-area models for smart memory building blocks (i.e., library of BA+ and standard cells). Using all of these inputs, the frontend rapidly explores the design space using BA+ macro-models and identifies candidate smart memory design points that could potentially meet users' architectural, physical and electrical specifications. More details about the SMSF frontend is found in a companion thesis in [47].

### 4.5.2 SMSF Backend

The generic SMSF backend explores the physical design space and creates an implementation of a smart memory design point that best meets specification. During implementation, the smart memory RTL, generated from SMSF frontend, is functionally verified through simulation and instantiates BA+ blocks that are controlled by the logic function described behaviorally. Next, this RTL is synthesized in a physically aware manner to a gate level netlist consisting of standard cell logic and BA+ tiles. This gate level netlist is then taken through a parameterized floorplanning step that is capable of implementing any design point identified by the SMSF frontend, for a given application. The application-specific parameterized floorplan captures the relative placement of macro blocks for every allowable design point. Our parameterized floorplanner tool uses the following inputs to create the floorplan for a specific design point: a) Synthesized area of the gate-level-netlist; b) Embedded memory partitioning parameters from the SMSF frontend RTL, such as, number of embedded memory banks, number of BA+ per bank etc.; c) Desired physical aspect ratio; d) Power-performance constraints file; and e) Physical (.lef) and timing (.lib) abstracts for the smart memory building blocks. With the floorplan sized and BA+ macro blocks placed, the parameterized floorplanner generates an optimized power plan and pin placement. Following these floorplanning steps, the physical implementation engine continues with the typical steps of placement, clock optimization and routing, aided by SMSF-specific wrapper scripts.

## 4.6  1R/1W SRAM Synthesis

As multi-ported embedded memories are extensively used to increase the memory operations per second (MOPS) in SoCs [42], we demonstrate how SMSF can be customized to build a 1Read-1Write (1R-1W) SRAM synthesis engine (Figure 4.7). The engine is split into a frontend for design space exploration and a backend for automatic physical implementation.

### 4.6.1  Frontend

The goal of a 1R-1W SRAM synthesis engine's frontend (Figure 4.7) is to identify SRAM configurations that could achieve user specified aspect ratio requirements. To improve efficiency, SRAMs are typically partitioned into several banks and placed in a row/column configuration, as shown in Figure 4.8. An SRAM "bank xy", located at "row x" and "column y", contains several BA+ that horizontally abut to share global read and write bitlines. The SRAM configurations that meet the user specified aspect ratio are identified by exhaustively iterating through the library of BA+ and their allowed configurations, given their physical dimensions. Once identified, we use the Genesis framework to generate a synthesize-able RTL for the chosen configurations; i.e. #SRAM bank rows, #SRAM bank columns, #BA+ per bank, for exploring the physical implementation design space using the SMSF backend.

### 4.6.2  Backend

Figure 4.7 illustrates the SMSF backend for synthesizing 1R-1W SRAM. First, the SMSF backend subjects the RTL for the chosen SRAM configurations (that met the user's aspect ratio specification) to logic synthesis to identify the design point that best meets the user specified timing target. Next, a parameterized floorplan is created for the SRAM configuration. Specifically, #SRAM bank rows, #SRAM bank columns, #BA+ per SRAM bank and aspect ratio are all used to create a relative BA+ macro placement, power plan and SRAM pin placement. Following this, standard cell placement, clock tree synthesis and routing are carried out using technology specific scripts.

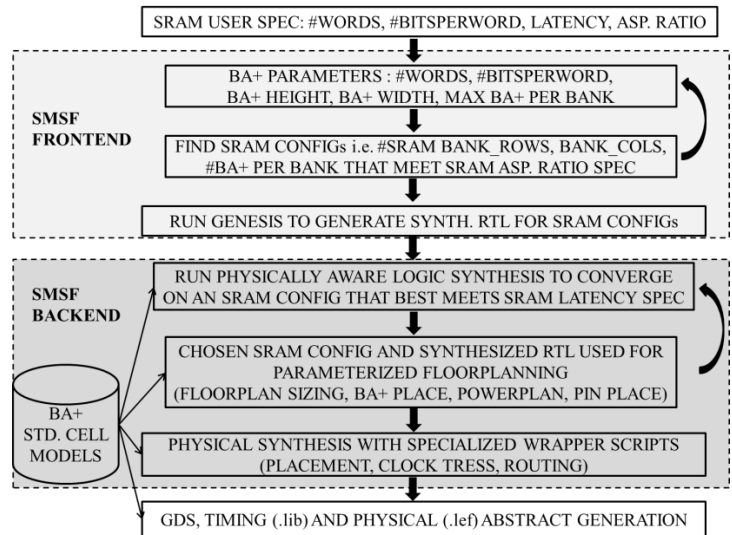

Figure 4.7 Customized SMSF for 1R-1W SRAM.

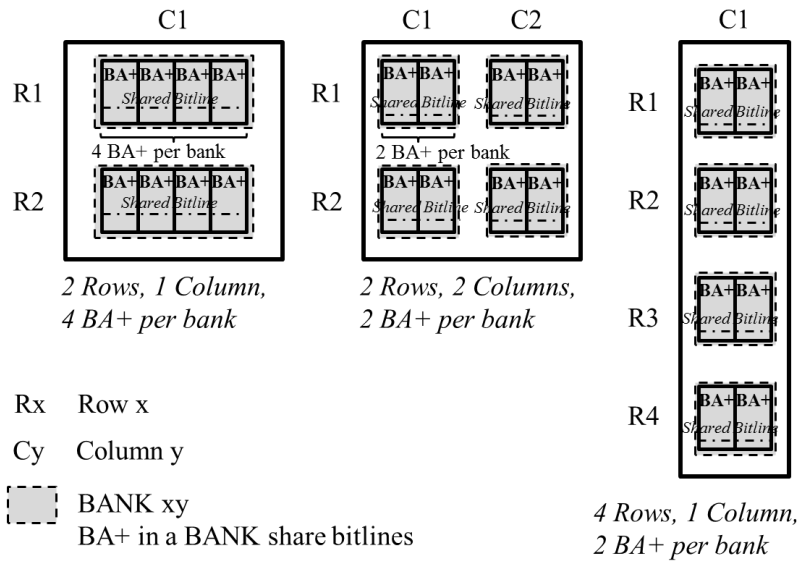

Figure 4.8 Example SRAM configuration used by SMSF customized to synthesize 1R-1W SRAMs.

Automating the generation of such an embedded memory IP greatly improves design turn-around times. The quality of the final physical implementation can be improved by aggressively customizing the SMSF for a given application at the cost of increased design time using standard physical synthesis optimization options such as placement constraints, Vt-optimization, power domain partitioning, etc. Therefore, SMSF can be customized for a specific application to create a powerful end-to-end tool for generating the smart memory hardware IP by rapidly evaluating design tradeoffs using standard cell logic and BA+ models.

### 4.6.3   Preliminary Results

As an illustration, we explore the design space for 256x8 and 256x32 SRAMs synthesized using a few different SRAM configurations, with two BA+s (32x8 and 32x32). It can be observed from performance-per-watt (GOPS per Watt) vs. area plots in Figure 4.9 that a designer could appropriately choose an SRAM design point that best meets his or her design goal. SMSF's ability to explore a vast design space, coupled with the capability to generate robust and efficient SRAM IP, improves design efficiency and design productivity.

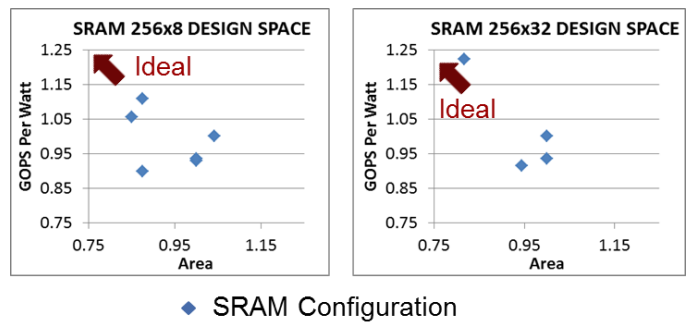

**Figure 4.9 SRAM design space exploration examples.**

### 4.7   Parallel Access SRAM Synthesis

We choose a high performance and energy efficient parallel access memory block used widely in imaging as an application example to evaluate the efficacy of smart memory synthesis as applied to sub-20 nm processes. It should be noted, however, that SMSF is customizable for a wide range of applications and benefits are not restricted to parallel access memories.

### 4.7.1 Parallel Access SRAM Application

Parallel access memory stores a 2D image pixel array and allows random access of pixels in a $2^a$ x $2^b$ block, in the vicinity of a given input address from among $2^m$ x $2^n$ pixels in one cycle. Figure 4.10 (a) show an example 2x2 (a=b=1) parallel access from a 32x32 (m=n=5) image. Parallel access memory is implemented traditionally (TM) using compiled SRAMs and standard cells by distributing pixel data in multiple SRAM banks to enable conflict-free access as shown in Figure 4.10(b). However, this traditional implementation does not exploit the address pattern commonality between the accessed pixels. Murachi et al. in [45] exploited the address pattern commonality and proposed an efficient application-specific SRAM with shared decoders (Figure 4.10(c)) that consumed significantly less power and area. Nevertheless, the high design cost of such a time intensive application-specific SRAM makes it unaffordable for SoC applications. Using SMSF we will demonstrate that a smart memory implementation (SM) of parallel access memories can achieve higher design efficiency comparable to a custom design while being designed using high productivity design flows.

### 4.7.2 Synthesis Engine

To demonstrate customizability and efficacy of SMSF we developed a smart parallel access memory synthesizer using a foundry 14 nm process design kit (PDK) and physical design tools. The synthesizer was created in a few hundred man hours using our tools and methodology, as opposed to thousands of man-hours consumed for creating a single custom design point as in [45]. The SMSF that we produced can generate designs with arbitrary image sizes and access window sizes by simply specifying the parallel access memory parameters as outlined by Zhu et al., in [47] where they have explored the applications, algorithms and the micro-architecture for the development of a synthesizable and parameterized smart parallel access SRAM that involves fine grained integration of logic and memory arrays (BA+).

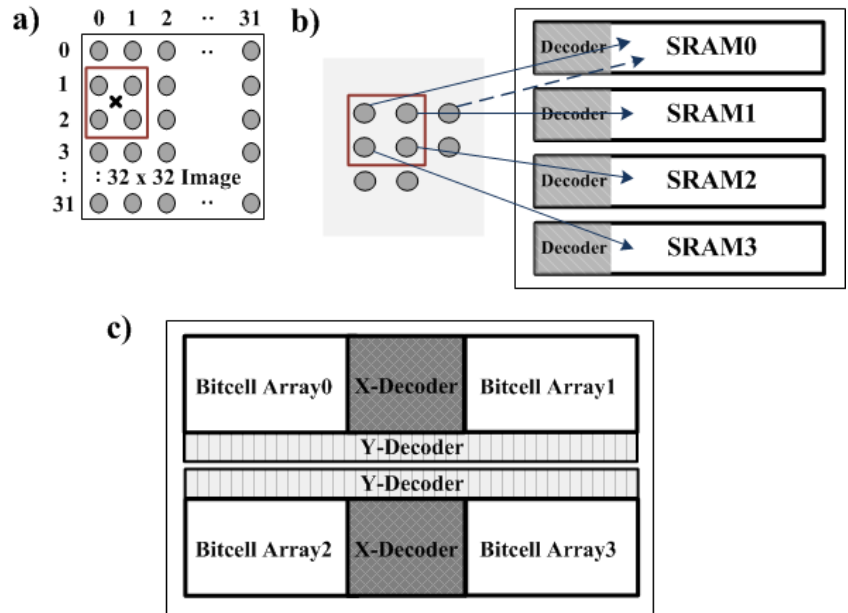

**Figure 4.10 (a) 2x2 parallel access from a 32x32 image (b) Micro-architecture of traditional SRAM based parallel access memory (TM) (c) Micro-architecture of smart parallel access memory (SM).**

### 4.7.3 Preliminary Synthesis Results

Figure 4.11 shows a physical design-space exploration of smart parallel access memory (SM). For ease of understanding, energy, delay and area for four design points pertaining to different parallel access window sizes, normalized w.r.t to a 2x2 window size, are presented. As expected, a larger parallel access 4x4 window size consumes more energy and area as compared to a smaller parallel access 2x2 window size, while providing higher image interpolation accuracy [47]. However, owing to pixel mapping into the embedded memory, 4x2 pixel access has similar hardware efficiency as 2x2 pixel access while supplying more pixels in a single clock cycle. Furthermore, Figure 4.11 highlights the energy-efficiency improvements observed in SM, compared to TM for different design points. It is noteworthy that the design space exploration presented here serves only as an example, and more user-desired analysis/exploration is possible with custom modifications to SMSF.

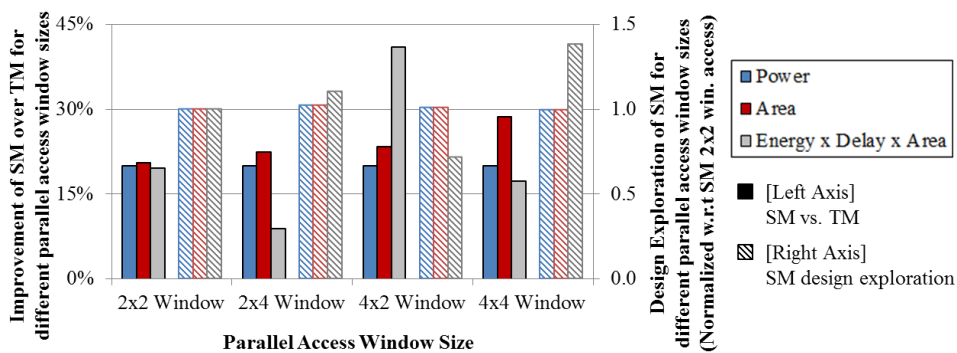

**Figure 4.11 Physical design space exploration results for parallel access SM using a smart memory synthesis framework created using a foundry 14 nm PDK. Comparison results between TM and SM is shown to demonstrate significant benefits.**

## 4.8 Summary

As leaf cell DTCO proves to be insufficient to drive affordable scaling, to best harness the features of a sub-20 nm CMOS process, we broaden the scope of DTCO to include micro-architecture and CAD along with circuits, layout and process technology. We subject the most significant component of a digital SoC, SRAM, to such a holistic DTCO, by synthesizing application-specific customization into the SRAM – smart SRAM. Smart SRAM blocks are built from robust primitives, augmented bitcell arrays (BA+) for storage and standard cells for peripheral and application-specific logic. While extremely restricted patterning degrades the efficiency of compiled SRAM blocks, we exploit this patterning impediment to tightly integrate bitcell arrays and standard cell logic, the two components that form the basis of area-efficient smart SRAM synthesis. Motivated by the possibility of robustly manufacturing smart SRAMs, we create a smart memory synthesis framework (SMSF). SMSF enables us to readily extend the benefits of synthesis to other application classes and embedded memories (DRAM, CAM etc.).

We demonstrate the adaptability of SMSF with two use cases, 1R-1W SRAM synthesis and parallel access SRAM synthesis. Preliminary simulation results indicate significant area and power benefits for smart SRAM implementations when compared with traditional IP based implementations.

# 5  Testchip and Experimental Results

One of the major goals of this dissertation is the deployment of the proposed design solutions onto a state-of-the-art process technology to evaluate and quantify their effectiveness. This exercise proved to be challenging but extremely useful as it clearly differentiated elegant concepts from usable engineering solutions. In this chapter we describe four different designs to evaluate the efficacy of the proposed approaches. The designs were fabricated in an IBM 14SOI electrical testsite and tested for function, performance and energy efficiency. The designs are listed below as follows:

- To evaluate the efficiency and manufacturability of the proposed standard cells, 10T_UniDir and 10T_BiDir, on silicon, we created ring oscillator test structures.
- We taped out a physically synthesized 10T_BiDir based 32-bit multiplier to test the readiness of the14SOI physical synthesis flow.
- To assess the feasibility and electrical benefits of synthesizing multi-port embedded memories, we synthesized and taped out an 8T-bitcell based 1R-1W 1KB SRAM along with an area-efficient compiled SRAM block serving as a baseline.
- To demonstrate the application-specific customizability of the proposed synthesis framework, we use the high performance parallel access SRAM used in imaging and video applications. This block was implemented in two ways: (a) traditional manner using compiled SRAM blocks and standard cell IP, and (b) using our proposed smart memory synthesis approach.

## 5.1  IBM 14SOI Design Infrastructure

### 5.1.1  Tools and Design Flows

Leaf cells were designed and laid out using IBM 14SOI process design kits (PDK). The PDK contained preliminary transistor and interconnect models for circuit simulation, and rule decks for physical verification. Block designs were physically synthesized from the leaf cells using the tools and flows shown in Figure 5.1. The physical synthesis flow was developed in collaboration with IBM and Cadence for 10T_BiDir standard cell library using the BEOL stack in Table 5.1.

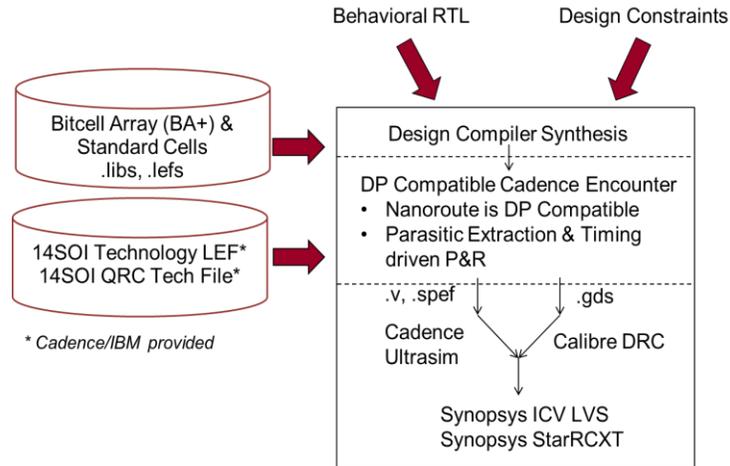

Figure 5.1 IBM 14SOI physical synthesis flow.

Table 5.1 BEOL stack supported for electrical testsite.

| Layer | BEOL Stack (Direction/Pitch) |
|---|---|
| PC | V/1y |
| M1 | H/1x and V/1y |
| M2 | H/1x |
| M3 | V/1x |
| M4 | H/1.5x and V/3x |
| M5 | V/1.5x and H/3x |

## 5.1.2  Design Integration into Testsite

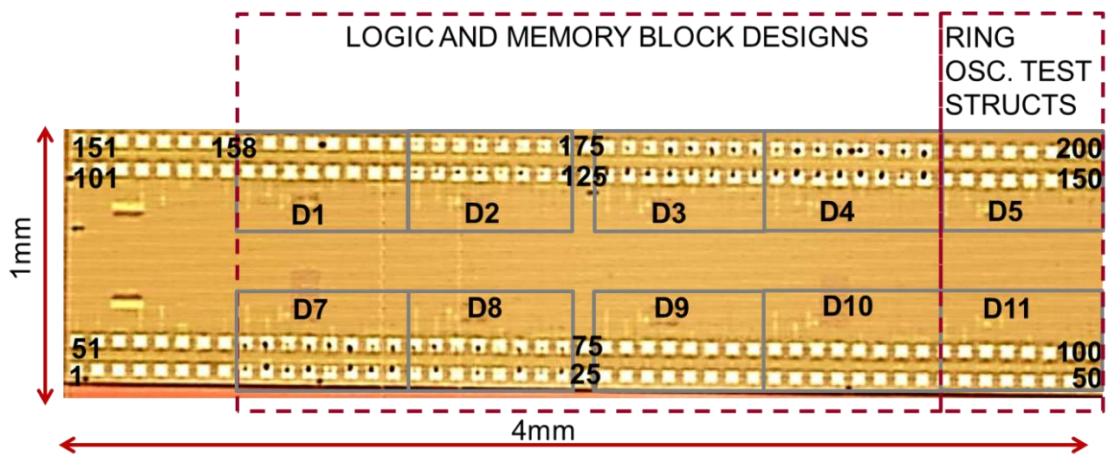

Figure 5.2 Micrograph from IBM 14SOI testsite containing our design blocks.

A micrograph of our design blocks integrated onto an IBM 14SOI electrical testsite is shown in Figure 5.2. All digital design blocks work in the GHz range of frequencies and have lots of high frequency inputs and outputs. While it is possible to test ring oscillators easily, integrating and testing other design blocks is not straight forward. To avoid expensive test setup and to minimize, if not eliminate, the number of high frequency inputs and outputs (I/O), all design blocks contain a shift/scan wrapper and a programmable clock generator, as shown in Figure 5.3. The programmable clock generator has 32 different frequency modes and also has dedicated supply (VDD_RO) for further frequency tuning, thereby generating frequencies from KHz to GHz. The shift/scan wrapper circumvents the use of high speed I/O by receiving inputs serially, buffering them on-chip and then feeding them to the design under test. After the design evaluates, output is captured by the same shift/ scan register and shifted out serially at low frequency (KHz or MHz).

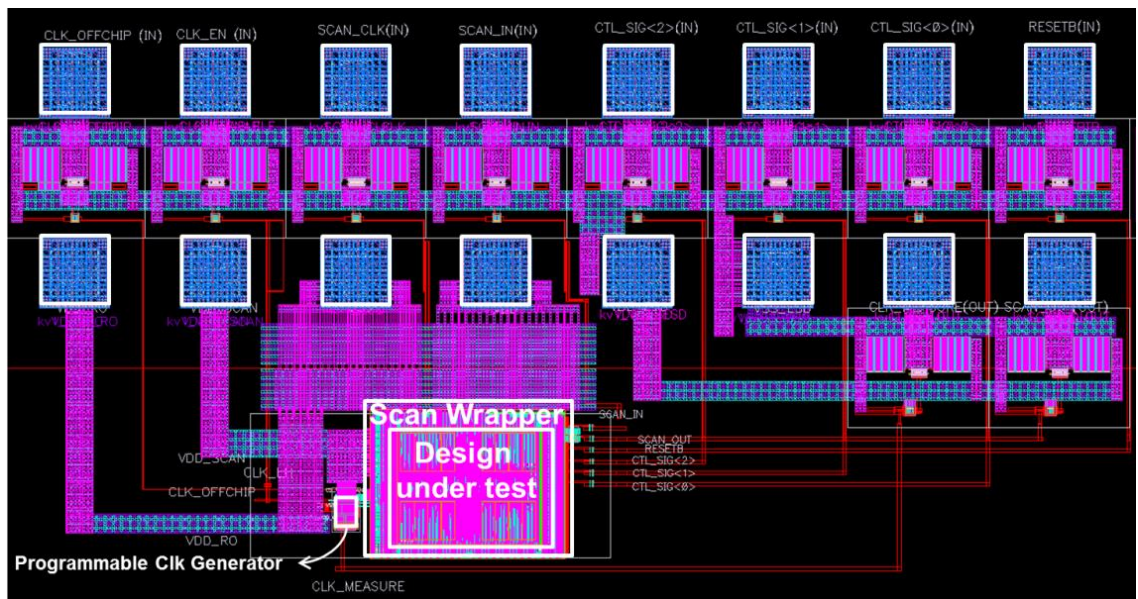

**Figure 5.3 Block design integration example.**

Logical and functional view of an example block design is shown in Figure 5.4 and Figure 5.5 respectively. At a high level, all design blocks except for ring oscillators are implemented with scan/shift I/O. Such a design contains 16 I/O of which 8 are inputs, 2 are outputs and 6 are power pins. To measure power, the design under test, multiplier, shown in Figure 5.4, has its own supply (VDD) isolated from ESD supply (VDD_ESD) and supply for scan/shift wrappers (VDD_SCAN). As noted previously, clock generator has its own dedicated supply (VDD_RO) to be able to generate a wide range of clock frequencies.

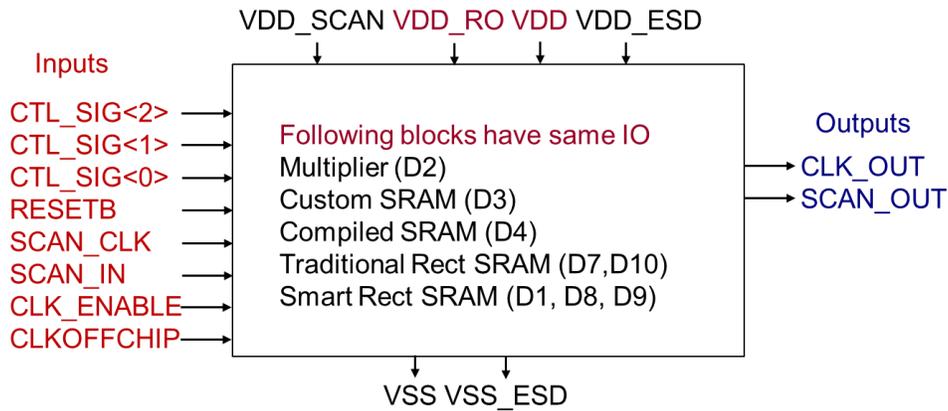

**Figure 5.4 Logical view of scan/shift based block design integrated on IBM 14SOI testsite.**

The shift/scan wrapper is controlled by 3 control signal bits (CTL_SIG<2:0>), where 5 (of the 8 possible states) control different operations as shown in Figure 5.5. A low frequency (few MHz) scan clock is used to synchronize I/O. The test sequence, as shown in Figure 5.6, starts with the RESETB signal driven to 0 and pulled back up to 1 to reset all sequential elements in the scan wrapper. Following that, a simple flush test is done to test the scan/shift wrapper and clock generator. After that the functional block is tested using vectors that can probe different functions in the design block. Using correctness of function at a particular clock frequency, observed using SCAN_OUT and CLK_OUT respectively, Shmoo plots can be created. To get better controllability during debug, we use dedicated ENABLE and CLKOFFCHIP pins to enable the clock generator and to drive the off-chip clock, respectively.

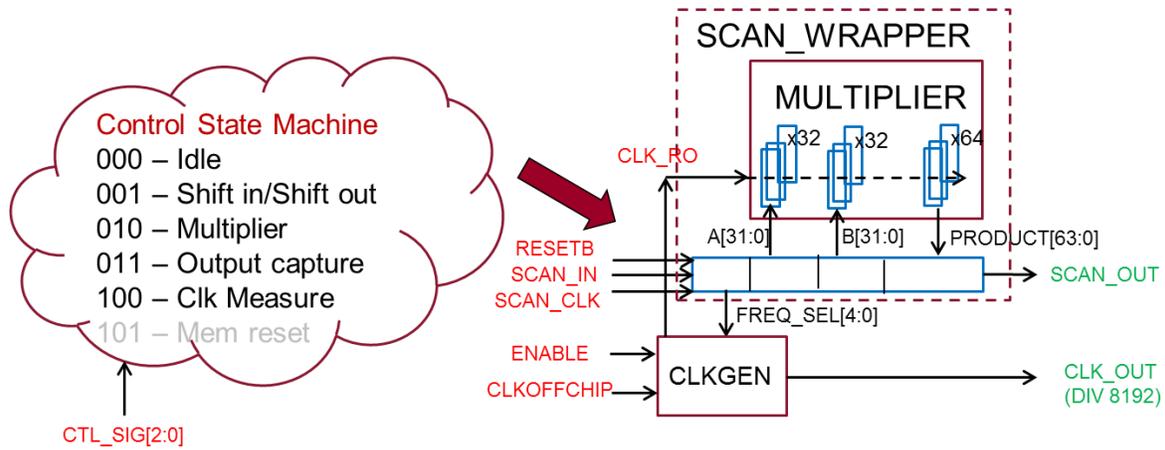
**Figure 5.5 Functional view of block design integrated on IBM 14SOI testsite.**

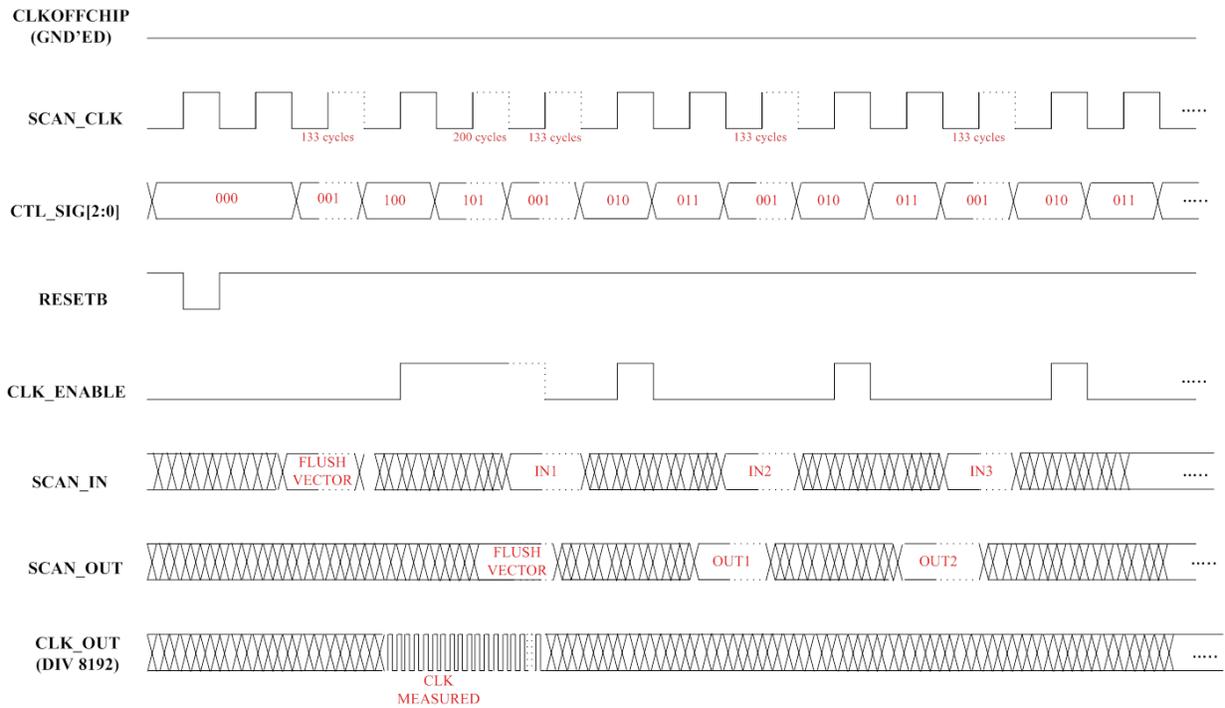

Setup violations are avoided by launching all inputs in negative edge of scan_clk and capturing them at the positive edge of scan_clk. For hold it's the viceversa.

**Figure 5.6 Timing diagram of shift/scan wrapper tested block design.**

## 5.2   IBM 14SOI Testing

The 14SOI design blocks were manufactured at the IBM Fishkill 300 mm fab. The designs were tested directly from 14SOI wafers with about 70 dies each. The wafers were tested in collaboration with engineers at IBM Fishkill using a custom wafer probe card, Advantest wafer tester and the Tel Prober shown in Figure 5.7.

For testing different design blocks in these wafers, we used the test flow shown in Figure 5.8. We generated test vectors using functional verification testbenches and custom scripts. Several test vectors were generated to excite different functions of a design block. As we were testing wafers from a pre-production electrical testsite, we first probed a design block in a die to check for correct function for a simple test vector at nominal voltage and a reasonable clock frequency. If the die fails the simple test for a design block, the die is not subjected to any further testing, significantly saving test time. If we observed the desired output, we then tested that die at different clock frequencies and different supply voltages using an elaborate suite of test vectors. For every die passing the suite of tests shown in Figure 5.8, values of clock frequency, leakage current, total current, supply voltage were appended spreadsheet (CSV file). After testing a design block for a few wafers, the CSV file was analyzed using custom scripts to generate supply voltage vs. frequency and supply voltage vs. power plots. This fully automated test-setup enabled us to test several wafers efficiently.

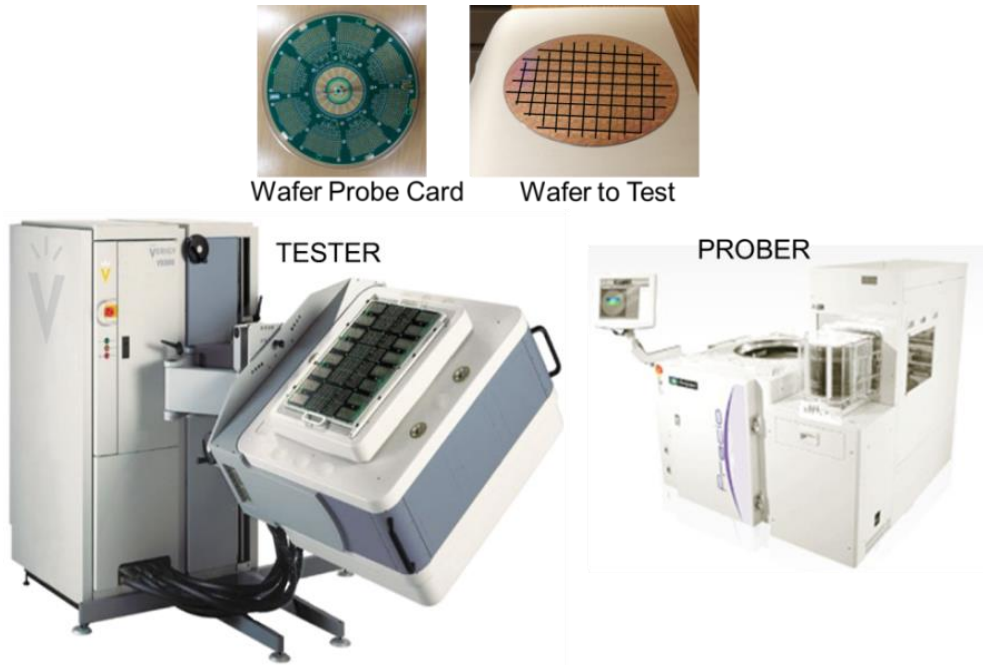

**Figure 5.7 Infrastructure used for IBM 14SOI wafer testing.**

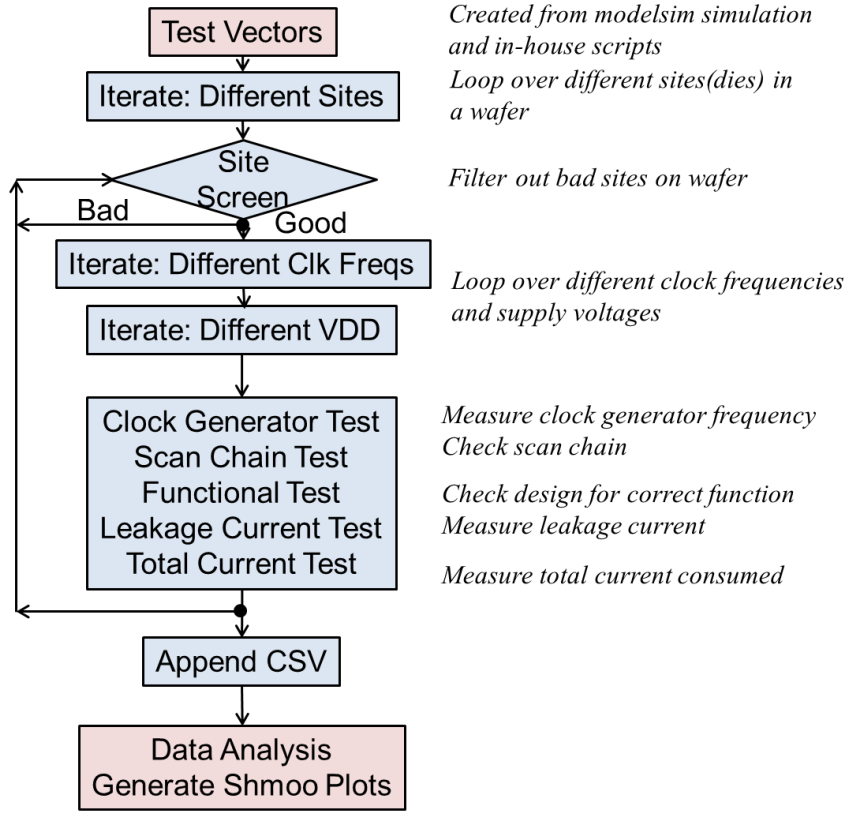

**Figure 5.8 14SOI wafer test flow.**

## 5.3 Standard Cell Based Ring Oscillator Test Structures

### 5.3.1 Design

In an effort to characterize power-performance of 10T_BiDir and 10T_UniDir standard cell architectures on silicon, we designed ring oscillator test structures as shown in Figure 5.9. The RO is turned ON by an ENABLE signal feeding a NAND stage, followed by 12 inverters. The NAND and inverters are drawn from the 10T_UniDir and 10T_BiDir libraries for the UniDir_RO and BiDir_RO, respectively. The RO frequency is divided down by a 14 bit divider.

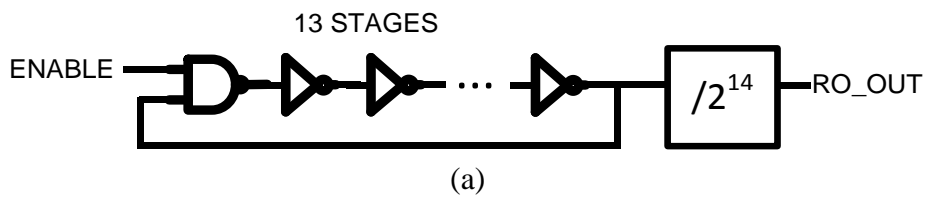

(a)

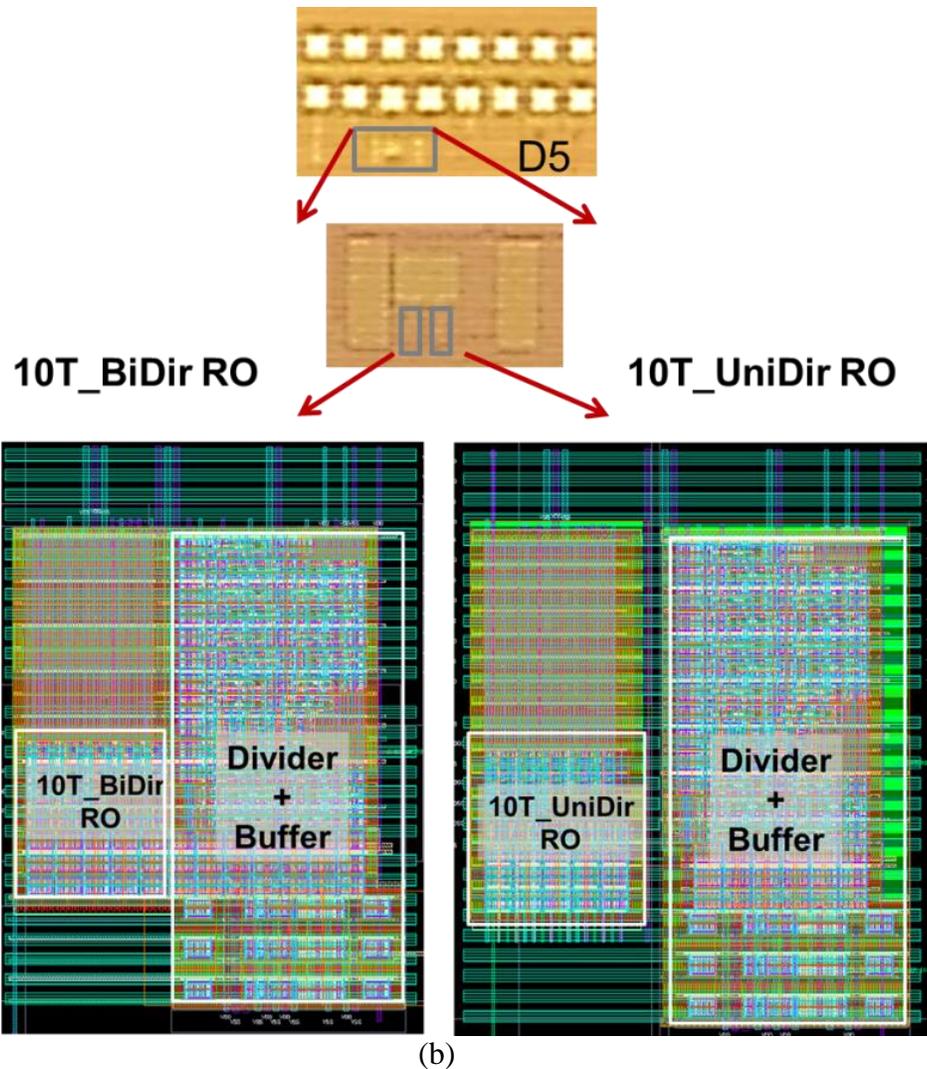

(b)
**Figure 5.9 (a) RO schematic, (b) Block micrograph and layout, of the RO test structures.**

## 5.3.2 Measurement Results

The ROs were subjected to tests at different operating conditions, namely, NOM+, NOM- and NOM. The measurement results are shown in Figure 5.10 and Figure 5.11, with mean and standard deviation of measurement ensemble plotted with markers and error bars respectively. The 10T_UniDir based RO was observed to be about 30% slower and 4X leakier than the 10T_BiDir based RO. Given that this trend was not seen in transistor level simulations with extracted parasitics, it could potentially be tracked down to two major differences in cell layouts, as seen in Figure 5.12:

- Gate connection near the edge of the cell in 10T_UniDir, as opposed to the gate connection in the center of the cell in 10T_BiDir, as shown in Figure 5.12 as (1). Such a non-conventional gate connection could result in increased gate resistance, slowing down 10T_UniDir RO. Additionally, if this gate connection is not manufactured reliably, then that could also result in increased gate leakage.
- NMOS and PMOS connection using local interconnect CA layer in 10T_UniDir as opposed to the M1 based connection seen in 10T_BiDir, as shown in Figure 5.12 as (2). CA based NMOS-PMOS connection running close to the gate of the two transistors could result in an increased gate-to-drain capacitance ($C_{gd}$), resulting in a slower speed.

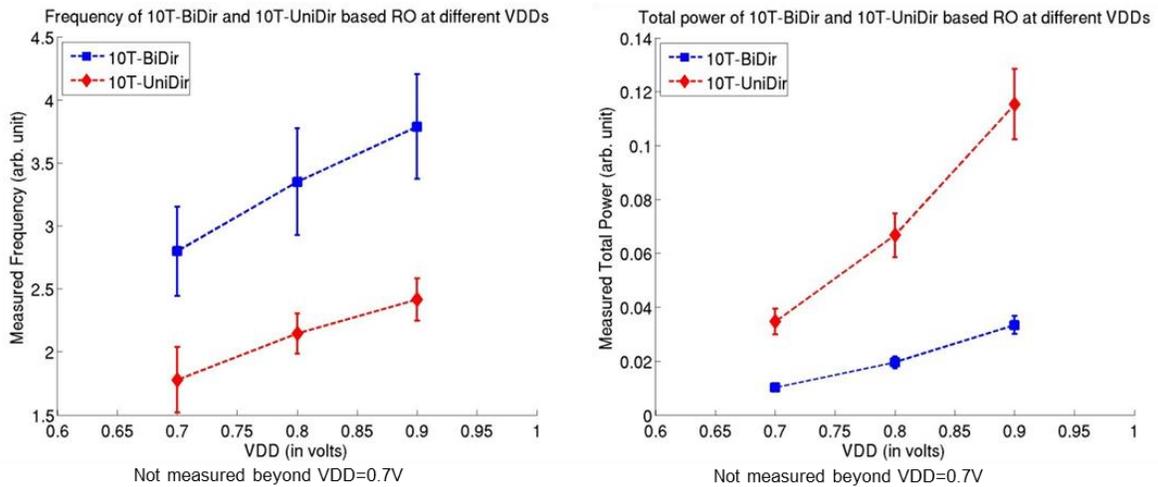

**Figure 5.10 Measurement results from 10T_UniDir and 10T_BiDir Ring Oscillators (RO). Plots show the mean as markers and standard deviation as error bars.**

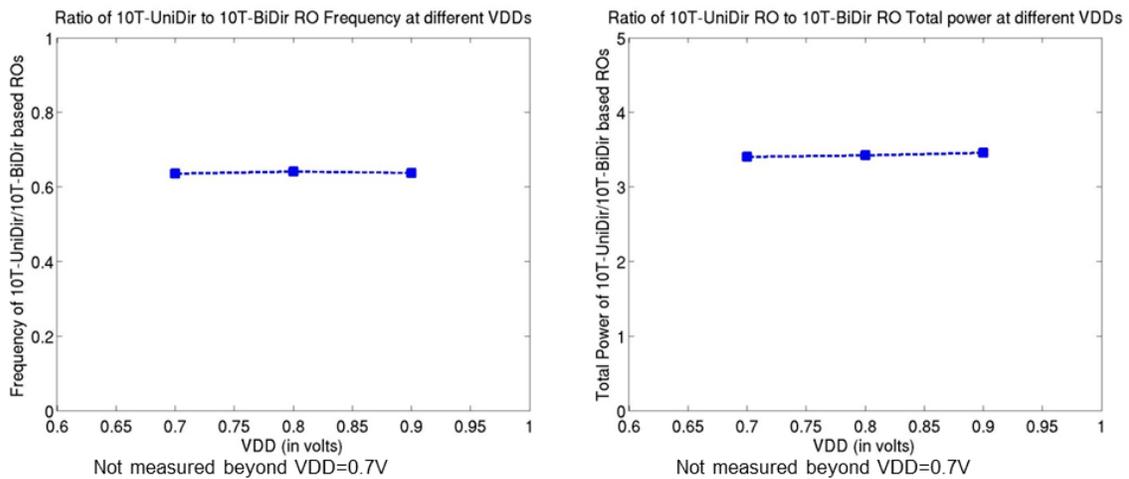

**Figure 5.11 Comparison of 10T_UniDir and 10T_BiDir ring oscillator(RO) for power and performance.**

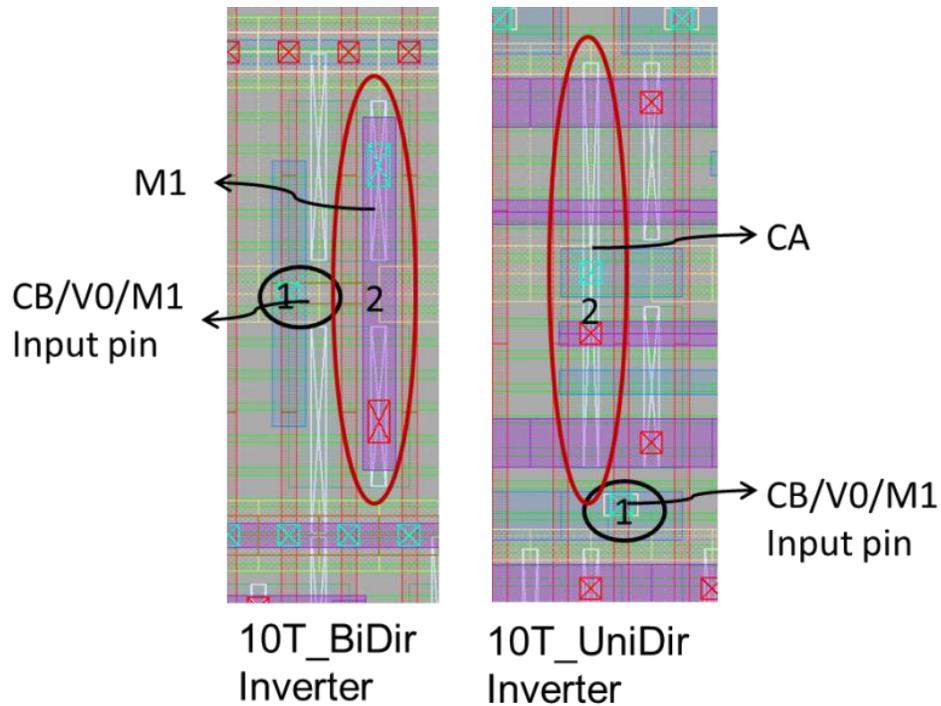

**Figure 5.12 10T_BiDir and 10T_UniDir inverter layouts highlighting the potential causes for the observed measurement results.**

Based on these results, compound grating-based standard cell (10T_BiDir) best trades off performance, area and manufacturability at the 14 nm technology node. Structured grating-based standard cell (10T_UniDir) is a very attractive alternative in terms of area, manufacturability and technology scalability, but suffers from sub-optimal performance and leakage at the 14 nm node.

## 5.4 Physical Synthesis Demonstration

### 5.4.1 Design

A 32-bit multiplier is implemented using the 10T_BiDir standard cell library and physical synthesis flow described in Section 5.1.1, as shown in Figure 5.13. The multiplier uses the Booth-Wallace topology. The multiplier is integrated into the testsite with shift/scan wrapper as described in Section 5.1.2.

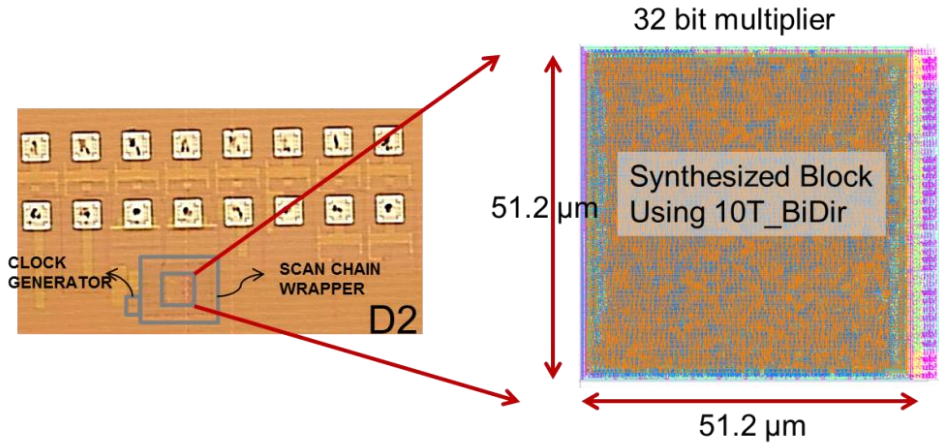

**Figure 5.13 32-bit multiplier micrograph and layout.**

### 5.4.2 Measurement Results

The multiplier was tested for functional correctness using different input vectors at different supply voltages. Measurement results from the 10T_BiDir based 32-bit multiplier are shown in Figure 5.14 and summarized in Table 5.2. Plots show the mean as markers and standard deviation as error bars. Measured data seems to be well in line with expectation. Owing to limitations of scan and test, electrical measurements could not be made below a supply voltage of 0.6V and frequencies about 6 GHz. As measurements above 6 Ghz were not possible, we could not test to see if the block works at frequencies higher than 6 GHz for higher supply voltages, which the flat trend seen for the frequency vs. supply voltage plot in Figure 5.14. This block functioning as per specification demonstrates the readiness and effectiveness of 14SOI physically synthesis flow.

**Table 5.2 32-bit multiplier summary of results.**

| Attributes | Results |
|---|---|
| Area | 2621.44 µm² |
| Gates | 4469 |
| Metals | 5 |
| GOPS per Watt | 3000 at 0.6V |
| Dynamic Range | 0.6-1V |

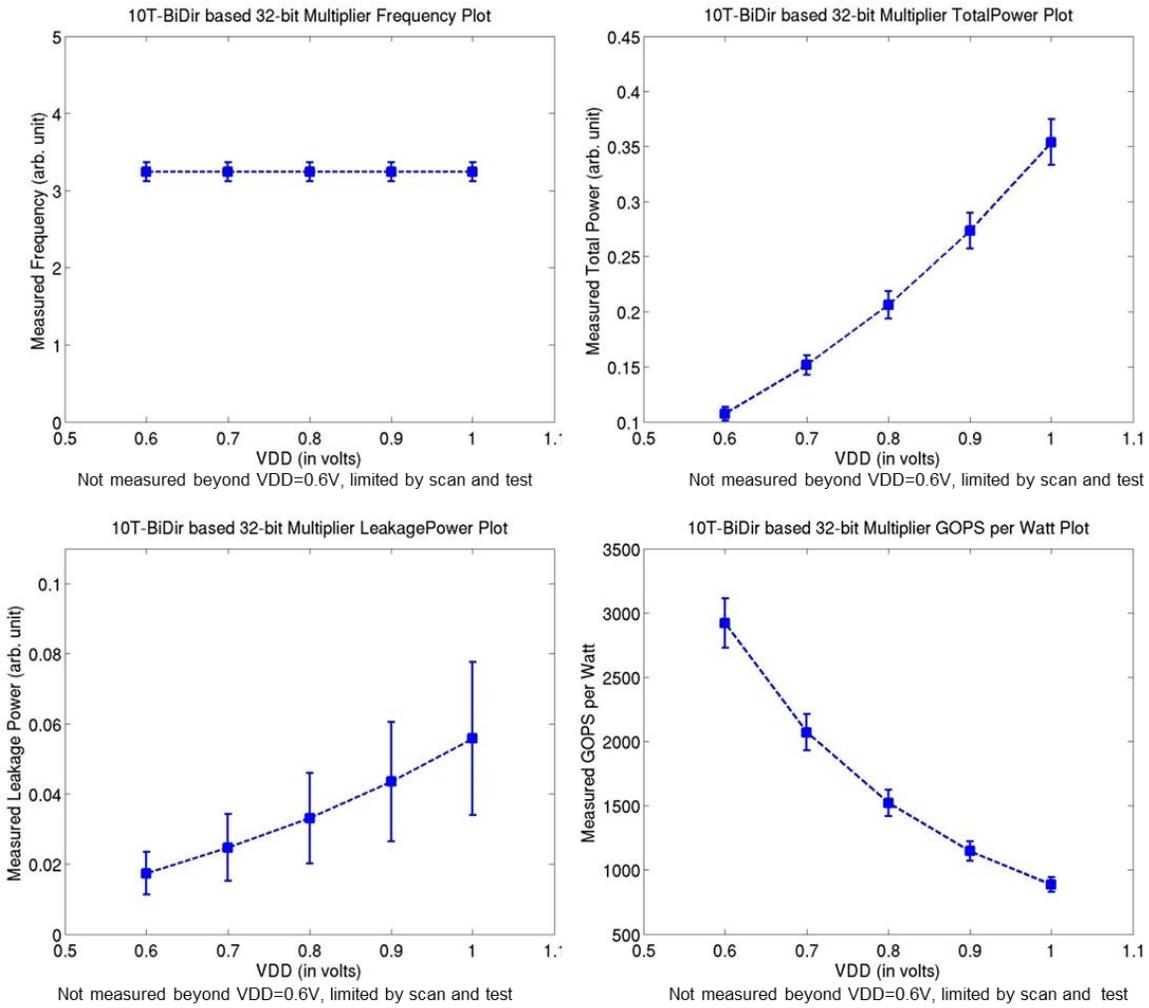

**Figure 5.14 Measurements from 10T-BiDir based 32-bit multiplier. Plots show the mean as markers and standard deviation as error bars.**

## 5.5 Smart SRAM Synthesis Demonstration

To demonstrate the efficiency and design productivity benefits offered by smart memory synthesis, we have synthesized a 1R-1W 1KB SRAM. Further, to demonstrate the ability to add application-specific features, we have synthesized a parallel access SRAM block that is used extensively in high performance imaging and video.

### 5.5.1 1R-1W SRAM Synthesis

#### 5.5.1.1 Design

The 1R-1W SRAM synthesis engine described in Section 4.6 is used to synthesize an instance of 256-word, 16-entry, 1Read-1Write (1R-1W) 1KB SRAM shown in Figure 5.15. As a baseline comparison, we also implemented a traditional design of a 1R-1W 256x16 1KB SRAM using the same augmented bitcell array circuits with compiled periphery. Leaf cells in the compiled periphery were designed using scalable circuit topologies while meeting 14SOI design rules.

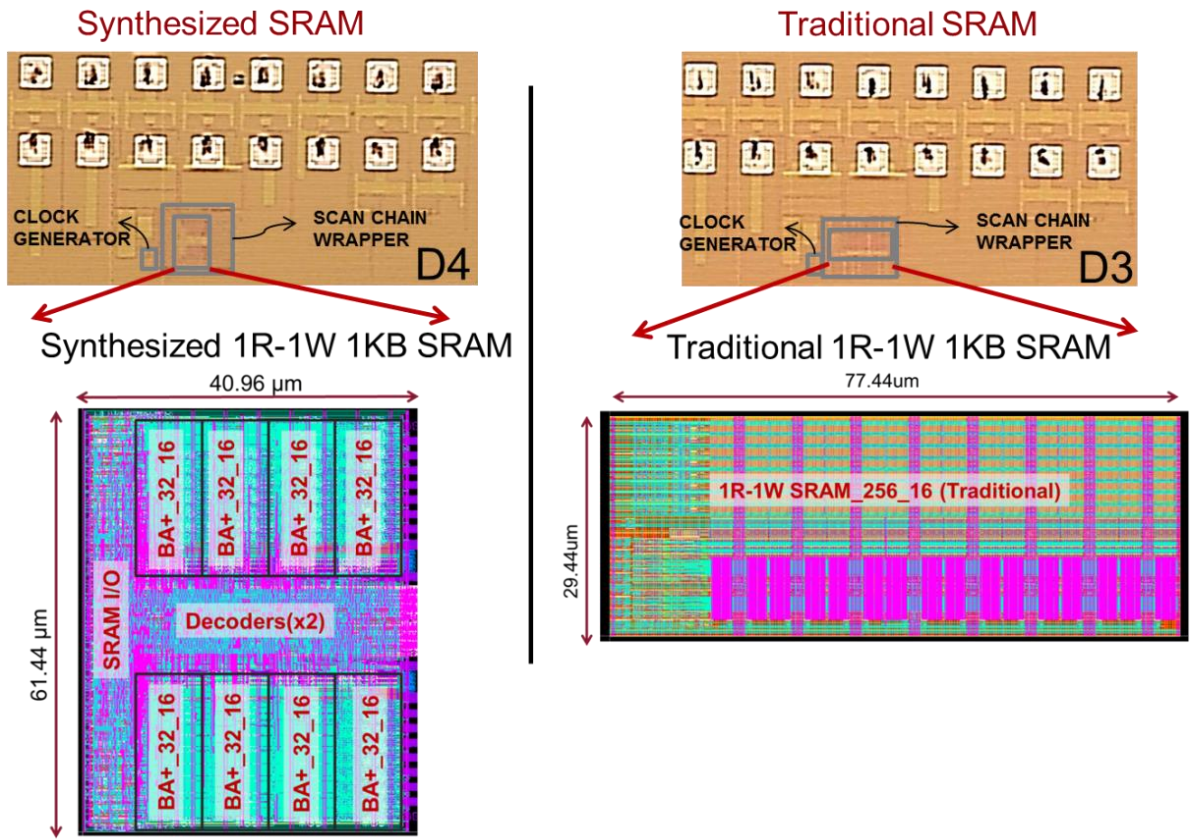

**Figure 5.15 Synthesized and traditional 1R-1W 1KB SRAM micrograph and layout.**

#### 5.5.1.2 Measurement Results

Both SRAM blocks were tested for functional correctness using several input vectors at different operating voltages using test flow shown in Figure 5.8. The measurements from the synthesized

SRAM and the traditional SRAM are shown in Figure 5.16, compared in Figure 5.17, and summarized in Table 5.3. We observe that the traditional SRAM, while consuming about 10% less area is slower than the synthesized SRAM by at least 1.5X. This speed degradation seen in traditional SRAM can be attributed to the inefficiency of patterning restricted leaf cells used by SRAM compilers (Section 3.5.1). Synthesized SRAM operating at higher frequency also consumes more power than traditional SRAM. However, performance-per-watt (Giga operations per second per watt i.e. GOPS per W) of the synthesized SRAM is still slightly better than traditional SRAM, especially at lower supply voltages. Both SRAMs demonstrate a moderate dynamic range with a $V_{min}$ of 0.6V, which is limited primarily by the capabilities of our scan and test system. This demonstration proves the ability of the SRAM synthesis engine to synthesize high performance and energy efficient embedded memories. Synthesized SRAMs are generated from robust primitives containing a small number of unique constructs, thereby making them more manufacturable. While SRAM synthesis can generate efficient SRAMs more productively than SRAM compilers, these synthesized "dumb" SRAMs could be made even more area and energy efficient by customizing them to their target application, as demonstrated next.

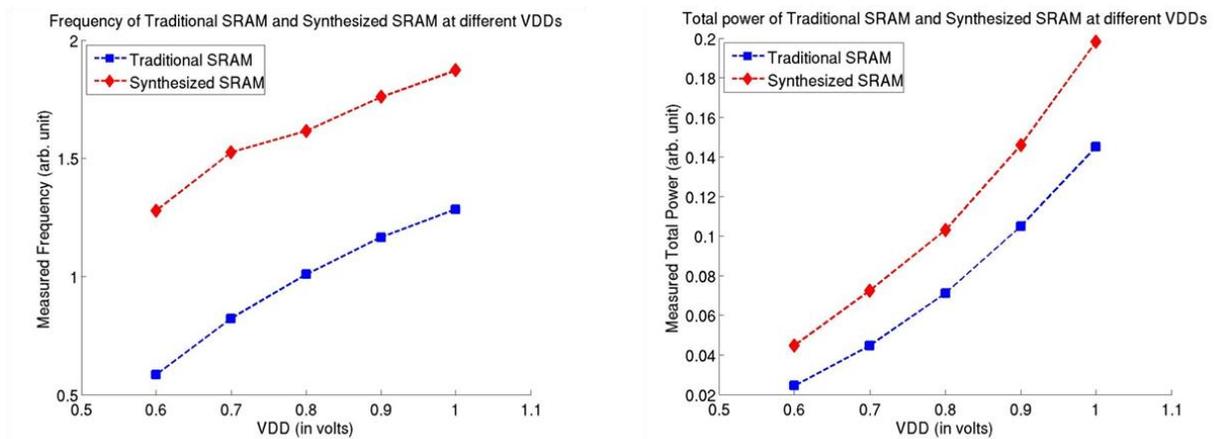

**Figure 5.16 Synthesized SRAM and traditional SRAM measurement results. Plots show the mean as markers.**

Table 5.3 Measurement summary table comparing Traditional and Synthesized SRAM.

| Attributes | Traditional SRAM | Synthesized SRAM |
|---|---|---|
| Area | 2279.8 µm$^2$ | 2516.58 µm$^2$ |
| Frequency (Norm) | 1.0 | 1.5X -2X |
| Energy per oper (Norm) | 1.0 | 0.85X-0.92X |
| Peak GOPS per Watt | 2400 at 0.6V | 2750 at 0.6V |
| Dynamic Range | 0.6-1V | 0.6-1V |

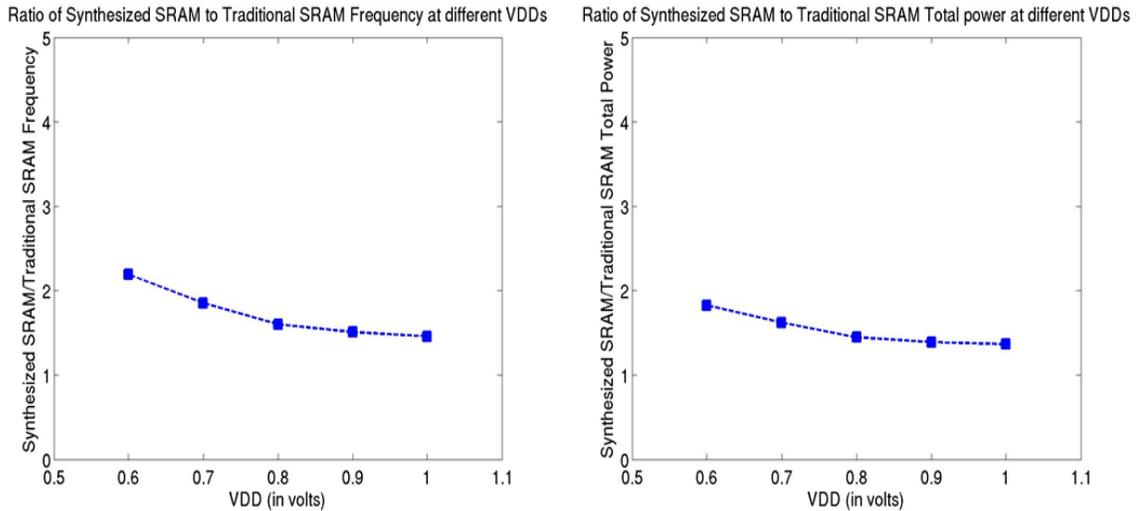

Figure 5.17 Comparison of synthesized SRAM and traditional SRAM for power and performance.

### 5.5.2 Parallel Access SRAM Synthesis

#### 5.5.2.1 Design

We evaluate the efficiency and customizability of the smart memory synthesis framework by implementing one specific design point of a smart parallel access SRAM (Smart PA SRAM) featuring a 2x2 block access from a 32x32-size image, and also a traditional parallel access SRAM (Traditional PA SRAM) with the exact same function for comparison. Both implementations used the same augmented bitcell arrays and were designed to meet 14SOI design rules. Block micrograph and taped out layouts are shown in Figure 5.18.

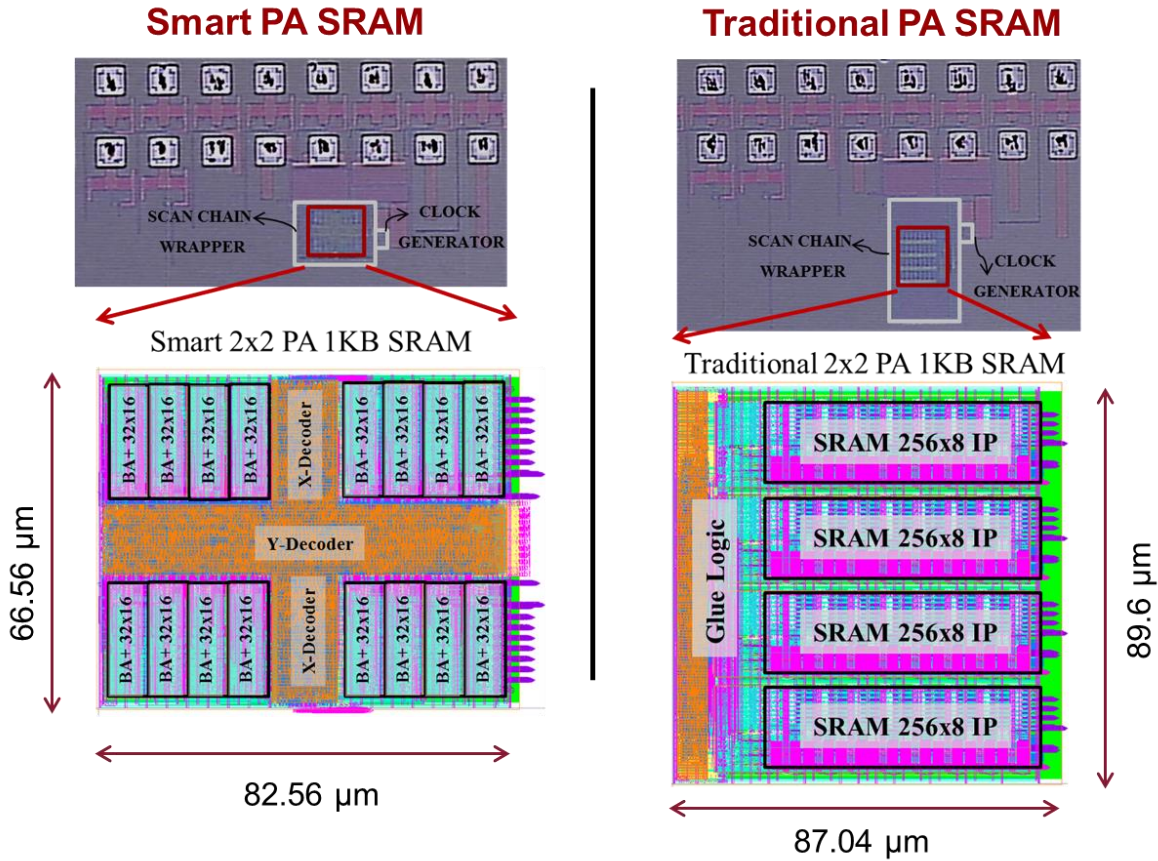

**Figure 5.18 Smart Parallel Access (PA) SRAM and Traditional Parallel Access (PA) SRAM micrographs and layouts.**

### 5.5.2.2 Measurement Results

Both PA SRAM designs were tested for functional correctness using different test vectors at different operating voltages. Measurements from smart PA SRAM (SM) and traditional PA SRAM (TM) are shown in Figure 5.19 and compared in Figure 5.20. Frequency measurements indicate that SM operates 2X to 4X faster than TM, while consuming slightly higher power. The smart 2x2 parallel access 1KB SRAM block also has an impressive performance per watt (GOPS per watt) of 2000 at 0.6V. While SM has a 50% better performance-per-watt compared to TM, it also consumes 25% less area (Table 5.4). SM achieves these improvements primarily by exploiting address pattern commonality inherent in the parallel access SRAM application. Customizing the SRAM architecture in an application-specific manner to support specialized merged X and Y decoders enables the SM block to be more efficient than TM. SM and TM both have a reasonable dynamic range of 0.5V with a $V_{min}$ of 0.6V, primarily limited by the scan and test system.

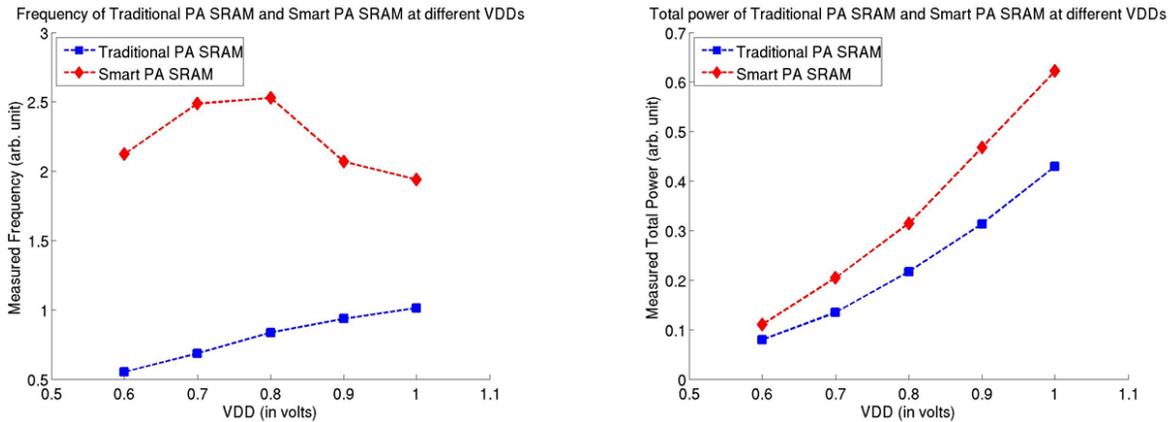

**Figure 5.19 Smart PA SRAM and Traditional PA SRAM measurements.**

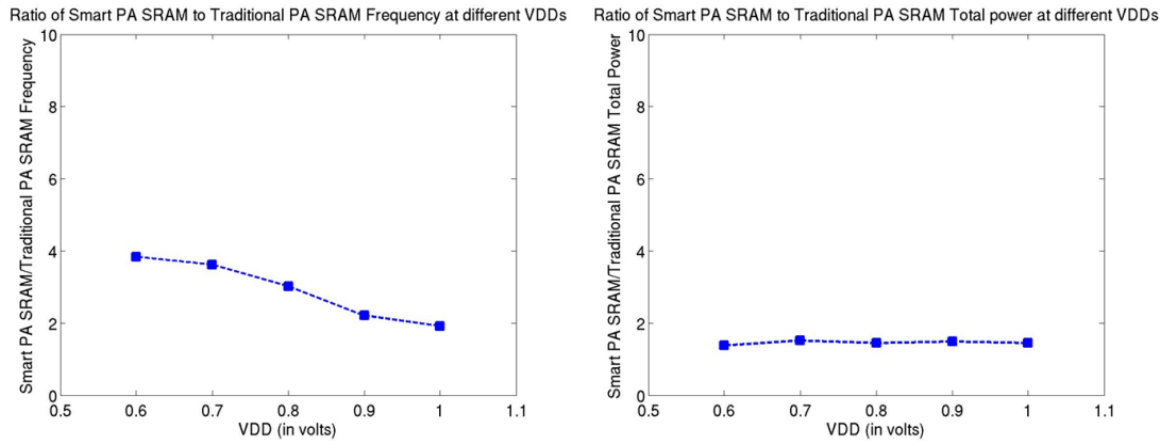

**Figure 5.20 Comparison of Smart PA SRAM and Traditional PA SRAM for power and performance.**

Fully functional and superior SM hardware:

- Quantifies the advantage of synthesizing application-specific customization into embedded memory using SMSF.
- Demonstrates the efficiency and readiness of all components in the 14SOI design infrastructure created by us, namely, 10T_BiDir standard cell, 8T BA+ blocks, physical synthesis flow etc.
- Illustrates improved manufacturability of synthesized SRAMs as they are generated from robust primitives that are built from a small set of unique pattern constructs.
- Ease of extensibility of SMSF to different application classes.

This design experiment motivates us to adapt more applications to SMSF to scale embedded memories in an efficient and cost effective manner. Zhu et al. extend SMSF to a few applications with promising benefits, as described in [47].

Table 5.4 Measurement summary table comparing Smart 2x2 PA and Traditional 2x2 PA.

| Attributes | Traditional 2x2 PA | Smart 2x2 PA |
| --- | --- | --- |
| Area | 7798.78 µm$^2$ | 5495.19 µm$^2$ |
| Frequency (Norm) | 1.0 | 2X - 4X |
| Avg. Energy per op (Norm) | 1.0 | 0.5X |
| Peak GOPS per Watt | 650 at 0.6V | 2000 at 0.6V |
| Dynamic Range | 0.6-1V | 0.6-1V |

## 5.6 Summary

To evaluate the effectiveness of the proposed approaches, we designed, fabricated and tested several experimental blocks in a state-of-the-art IBM 14SOI process. Measurement results indicate that the proposed techniques enable us to improve design efficiency, decrease manufacturing cost, improve manufacturability and decrease turnaround time, therefore driving affordable scaling into the future. A summary of experimental results is listed below:

1) In our first experiment we evaluated 10T_UniDir and 10T_BiDir for performance and power dissipation. Both standard cells were observed to have a similar area foot-print but the 10T_UniDir demonstrated better manufacturing scalability to smaller technology nodes. Measurements from 10T_UniDir and 10T_BiDir based ring-oscillator (RO) test structures showed that 10T_BiDir has a performance advantage of about 30% over the 10T_UniDir, while simultaneously leaking 4 times less current.

2) To test the efficacy of the 14SOI physical synthesis flow, we taped out a physically synthesized 10T_BiDir based 32-bit multiplier. Functional testing showed the block to perform in line with expectations. This experiment shows the readiness of the 14SOI physical synthesis flow to generate area and energy efficient design blocks using 10T_BiDir standard cells. The synthesized blocks contain a small set of unique pattern constructs that makes them more cost effective to manufacture and amenable to scale.

3) In our next experiment, we evaluated the manufacturing feasibility and electrical benefits of synthesizing multi-port embedded memories. An 8T-bitcell based 1R-1W high performance 256-word, 16-entry, 1KB SRAM was synthesized and taped out along with an area-efficient, traditionally designed SRAM serving as a baseline, in the IBM 14SOI testsite. Electrical measurements revealed that synthesized SRAM delivers 2X better performance while consuming 10% more area.

4) In our final experiment, we evaluated the efficacy of our proposed approaches using a real SoC sub-block, namely the parallel access SRAM. The goal of this exercise was to integrate the different pieces of design infrastructure created as part of this dissertation, namely, compound grating-based standard cells, synthesized smart SRAMs and 14SOI physical synthesis flow, to create a high performance, energy efficient and manufacturable design at 14 nm and beyond. The parallel access SRAM block was implemented in two ways: (a) traditional manner using SRAM and standard cell IP serving as baseline, and (b) using our proposed smart memory synthesis approach. Electrical testing revealed that both design blocks were fully functional. The smart parallel access SRAM outclassed the traditional parallel access SRAM as it was 25% more area efficient, two-four times faster, while having a 50% better performance-per-watt. Synthesized implementation for such application-specific SRAMs also dramatically reduces design turnaround time. Furthermore, synthesized SRAMs are more amenable to manufacturing as they are built from a small set of unique pattern constructs. Therefore, these results emphasize that such a holistic DTCO process is essential to drive affordable scaling in the future.

# 6  Conclusion and Future Work

As current design techniques fail to address the daunting challenge of affordable SoC design and manufacturing, there is an imminent need for a solution that improves design efficiency, reduces manufacturing cost, improves manufacturability and decreases design turnaround time. To this end we undertook circuit, layout and technology co-optimization for two fundamental leaf cells, namely standard cells and SRAM bitcells. Leaf cell DTCO improved manufacturability by minimizing the number of unique pattern constructs, lowered manufacturing cost by allowing the use of cost-effective patterning techniques and improved design efficiency by achieving leaf cell area scaling of ~60% (anticipated 40%-50% [12]), compared to similar 32 nm cells. However, leaf cell DTCO was inadequate to meet Moore's area scaling requirement (75% from 32 nm to 14 nm), imperative for affordable scaling.

To tackle this formidable problem, we needed to exploit the features of a sub-20 nm process, while working past its impediments. To best exploit the challenges of sub-20 nm CMOS, we have to broaden the scope of design technology co-optimization to include micro-architecture and CAD, along with circuit, layout and process technology. We applied such a holistic DTCO to different components of a modern digital SoC design, to observe that it dramatically improves design efficiency while decreasing manufacturing cost and improving manufacturability, as summarized in Figure 6.1.

Particularly, subjecting the most significant component of an SoC, SRAMs, to holistic DTCO enabled us to synthesize SRAMs with application-specific customizations, from robust bitcell arrays and standard cell primitives. Smart memory synthesis drives affordable scaling in three ways. First, synthesis enables a designer to customize the internals of an SRAM to suit application needs, thereby significantly improving area and energy efficiency. Second, synthesizing such smart SRAM blocks from robust bitcell arrays and standard cell primitives that are built from a small set of constructs guarantees improved manufacturability. Third, smart memories are amenable to automatic generation through simple physical synthesis flows, thereby dramatically reducing design turnaround time while also enabling integration into mainstream SoC design flows.

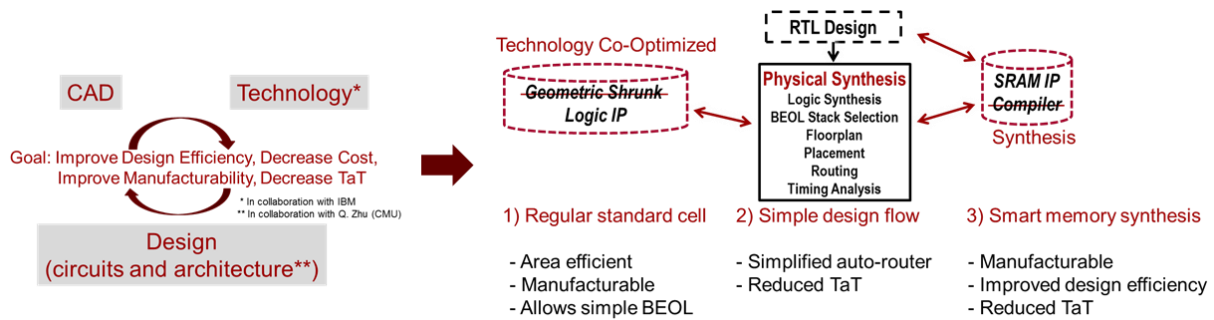

**Figure 6.1 Holistic DTCO - Concept to Practice.**

Applying holistic DTCO to logic blocks enabled us to build an efficient physical synthesis flow that used a simplified BEOL stack, gridded auto-router and pin access optimized standard cell library. Such an approach enabled us to reduce design turnaround time while improving manufacturability as the physically synthesized blocks were built from a small set of layout pattern constructs.

To evaluate the efficacy of the proposed methods, we designed, fabricated and tested an SoC sub-block, namely, parallel access SRAM, in a state-of-the-art IBM 14SOI process. Measurements from the smart parallel access SRAM block indicate a 25% improvement in area efficiency and 50% improvement in performance-per-watt, compared to a baseline design built using SRAM and standard cell IP blocks. The smart parallel access SRAM block also scaled by 74% compared to a traditional parallel access SRAM block in 32 nm node, almost meeting the ideal 75% Moore's area scaling requirement.

These experimental results lead us to conclude that such holistic DTCO is essential to scale in a cost-effective manner below 20 nm CMOS, as such an approach enables improved design efficiency, decreased manufacturing cost, improved manufacturability and decreased turnaround times. To extend the benefits of smart memory synthesis to a wide class of special functions seen in modern SoCs, we have created the smart memory synthesis framework (SMSF). Zhu et al. have examined several applications where SMSF could be applied and preliminary results look promising [47].

In conclusion, this dissertation has proposed and evaluated an alternative path to affordable scaling, readily deployable at the 14 nm node. The proposed design and CAD solutions are also scalable down to 10 nm and 7 nm node, as they build from robust and manufacturable primitives whose scalability has been demonstrated in [48]. Conclusions from this dissertation open up new interesting research problems for future work, with a few listed below:

- SMSF emerges as a promising alternative to conventional SRAM IP compilation. Its wider applicability to other SoC blocks can greatly improve the overall design efficiency, quality and productivity. A comparison between the efficiency of an SoC built using compiled SRAM blocks and the same SoC designed using SMSF synthesized SRAM blocks could best demonstrate the readiness and benefits of SMSF.
- Automating augmented bitcell array schematic, layout and abstract generation can further improve the productivity of SMSF.
- Compound grating-based standard cells (10T_BiDir) emerge as the preferred choice at 14SOI, owing to their superior performance and leakage. However, since they are not the most amenable to scale, it is critical to explore methods to improve up on the efficiency of the more scalable structured grating-based standard cells (10T_UniDir).